\documentclass[a4paper,UKenglish,cleveref, autoref, thm-restate]{lipics-v2021}
\pdfoutput=1 %uncomment to ensure pdflatex processing (mandatatory e.g. to submit to arXiv)
\hideLIPIcs  %uncomment to remove references to LIPIcs series (logo, DOI, ...), e.g. when preparing a pre-final version to be uploaded to arXiv or another public repository
\nolinenumbers

\graphicspath{{./graphics/}}%helpful if your graphic files are in another directory

\title{Scroll nets} %TODO Please add

\author{Pablo Donato}{Charles University, Czechia}{pablo.donato@tuta.io}{0000-0001-7883-6754}{}{}%TODO mandatory, please use full name; only 1 author per \author macro; first two parameters are mandatory, other parameters can be empty. Please provide at least the name of the affiliation and the country. The full address is optional. Use additional curly braces to indicate the correct name splitting when the last name consists of multiple name parts.

\authorrunning{Pablo Donato} %TODO mandatory. First: Use abbreviated first/middle names. Second (only in severe cases): Use first author plus 'et al.'

\Copyright{Pablo Donato} %TODO mandatory, please use full first names. LIPIcs license is "CC-BY";  http://creativecommons.org/licenses/by/3.0/

 %TODO mandatory: Please choose ACM 2012 classifications from https://dl.acm.org/ccs/ccs_flat.cfm 


\category{} %optional, e.g. invited paper

\relatedversion{} %optional, e.g. full version hosted on arXiv, HAL, or other respository/website
\EventEditors{John Q. Open and Joan R. Access}
\EventNoEds{2}
\EventLongTitle{42nd Conference on Very Important Topics (CVIT 2016)}
\EventShortTitle{CVIT 2016}
\EventAcronym{CVIT}
\EventYear{2016}
\EventDate{December 24--27, 2016}
\EventLocation{Little Whinging, United Kingdom}
\EventLogo{}
\SeriesVolume{42}
\ArticleNo{23}
\usepackage[utf8]{inputenc}
\usepackage[T1]{fontenc}
\usepackage{stmaryrd}
\usepackage{mathpartir}
\usepackage{prftree}
\usepackage{tikz}
\usepackage{tikzit}
\usepackage{textcomp}
\usepackage{framed}
\usepackage{adforn}
\usepackage{bussproofs}
\usepackage{xcolor}
\usepackage{csquotes}
\usepackage{mathtools}
\usepackage[notion,electronic]{knowledge}
\input{notions.kl}
\knowledgeconfigure{quotation}

\usetikzlibrary{calc}
\usetikzlibrary{decorations.pathmorphing,decorations.pathreplacing,decorations.shapes}
\usetikzlibrary{arrows.meta}
\tikzstyle{pos text}=[fill=none, draw=none, text=pfg, inner sep=1pt]
\tikzstyle{neg text}=[fill=none, draw=none, text=nfg, tikzit fill={rgb,255: red,191; green,191; blue,191}, tikzit draw={rgb,255: red,191; green,191; blue,191}, inner sep=1pt]
\tikzstyle{source}=[fill=hypo, draw=none, shape=circle, tikzit fill={rgb,255: red,0; green,123; blue,255}, minimum size=4pt, inner sep=0pt]
\tikzstyle{target}=[fill=conc, draw=none, shape=circle, tikzit fill={rgb,255: red,236; green,26; blue,26}, minimum size=4pt, inner sep=0pt]
\tikzstyle{westhook}=[fill=white, draw=black, shape=diamond, tikzit shape=circle, anchor=west, inner sep=1pt]
\tikzstyle{westpred}=[fill=none, draw=none, shape=circle, anchor=west]
\tikzstyle{binder}=[fill=black, draw=black, shape=circle, inner sep=0pt, minimum size=3pt]
\tikzstyle{scroll anchor}=[fill={rgb,255: red,255; green,111; blue,0}, draw={rgb,255: red,255; green,111; blue,0}, shape=circle, inner sep=0pt, minimum size=3pt]
\tikzstyle{southbox}=[fill=none, draw=black, shape=rectangle, anchor=south]
\tikzstyle{northbox}=[fill=none, draw=black, shape=rectangle, anchor=north]
\tikzstyle{eastbox}=[fill=none, draw=black, shape=rectangle, anchor=east]
\tikzstyle{westbox}=[fill=none, draw=black, shape=rectangle, anchor=west]
\tikzstyle{box}=[fill=none, draw=black, shape=rectangle]
\tikzstyle{sep}=[fill=black, draw=black, shape=circle, inner sep=0pt, minimum size=3pt]

\tikzstyle{pos area}=[-, fill=pbg, draw=black, tikzit fill=white, tikzit draw=black]
\tikzstyle{neg area}=[-, fill=nbg, draw=black, tikzit fill={rgb,255: red,191; green,191; blue,191}, tikzit draw=black]
\tikzstyle{pistil}=[-, tikzit fill={rgb,255: red,255; green,243; blue,68}, tikzit draw=black, fill=pistil, draw=black]
\tikzstyle{justif}=[draw={rgb,255: red,0; green,210; blue,0}, ->]
\tikzstyle{teridentity}=[-, fill=black]
\tikzstyle{wire}=[-]
\tikzstyle{black arrow}=[->]
\tikzstyle{loop}=[->, distance=10mm]
\tikzstyle{attachment}=[-, draw={rgb,255: red,255; green,128; blue,0}, line width=1pt]

\newcommand{\tikzscale}[3]{
  \scalebox{#1}{
    \begingroup
      \tikzset{every picture/.style={scale=#2,baseline=(current bounding box.center)}}
      #3
    \endgroup}}
\renewcommand{\tikzfig}[3]{
  \tikzscale{#1}{#2}{\InputIfFileExists{./graphics/#3.tikz}{}{{\color{red}\colorbox{pink}{missingfile : #3}}}}}
\newcommand{\stkfig}[2]{\input{#1.tikz}{0.5}{#2}}

\definecolor{pbg}{HTML}{FFFFFF}
\definecolor{nbg}{HTML}{DFDFDE}
\definecolor{pfg}{HTML}{000000}
\definecolor{nfg}{HTML}{000000}

\newcommand{\includepdf}[2]{\vcenter{\hbox{\includegraphics[width=#1]{#2.pdf}}}}

\newcommand{\hrefl}[1]{\text{\raisebox{\depth}{\scalebox{-1}[1]{$#1$}}}}

\DeclareFontFamily{U}{BOONDOX-calo}{\skewchar\font=45 }
\DeclareFontShape{U}{BOONDOX-calo}{m}{n}{
  <-> s*[1.05] BOONDOX-r-calo}{}
\DeclareFontShape{U}{BOONDOX-calo}{b}{n}{
  <-> s*[1.05] BOONDOX-b-calo}{}
\DeclareMathAlphabet{\mathcalboondox}{U}{BOONDOX-calo}{m}{n}
\SetMathAlphabet{\mathcalboondox}{bold}{U}{BOONDOX-calo}{b}{n}
\DeclareMathAlphabet{\mathbcalboondox}{U}{BOONDOX-calo}{b}{n}

\makeatletter
\DeclareFontEncoding{LS1}{}{\noaccents@}
\makeatother
\makeatletter
\DeclareFontEncoding{LS2}{}{\noaccents@}
\makeatother
\DeclareFontSubstitution{LS1}{stix}{m}{n}
\DeclareFontSubstitution{LS2}{stix}{m}{n}

\DeclareSymbolFont{xsymbols3}{LS1}{stixbb}{m}{n}
\DeclareSymbolFont{xintegrals}{LS2}{stixcal}{m}{n}

\DeclareFontFamily{U}{mathx}{}
\DeclareFontShape{U}{mathx}{m}{n}{<-> mathx10}{}
\DeclareSymbolFont{mathx}{U}{mathx}{m}{n}

\DeclareMathSymbol{\rightwhitetriangle}{\mathrel}{mathx}{234}
\DeclareMathSymbol{\rightblacktriangle}{\mathrel}{mathx}{233}
\DeclareMathSymbol{\Coloneq}{\mathrel}{xsymbols3}{196}
\knowledgenewrobustcmd{\Ver}{\cmdkl{V}}
\knowledgenewrobustcmd{\pVer}[1]{\cmdkl{V^+_{#1}}}
\knowledgenewrobustcmd{\nVer}[1]{\cmdkl{V^-_{#1}}}
\newcommand{\NEdg}{\cmdkl{\rightarrowtriangle}}
\knowledgenewrobustcmd{\Edg}{\mathrel{\NEdg}}
\newrobustcmd{\XEdg}{\withkl{\kl[\Edg]}{\NEdg}}
\newcommand{\NtEdg}[1]{\cmdkl{\rightarrowtriangle^*_{#1}}}
\knowledgenewrobustcmd{\tEdg}[1]{\mathrel{\NtEdg{#1}}}
\newrobustcmd{\XtEdg}[1]{\withkl{\kl[\tEdg]}{\NtEdg{#1}}}
\knowledgenewrobustcmd{\Gra}{\cmdkl{G}}
\newcommand{\NAtt}{\cmdkl{\hrefl{\propto}}}
\knowledgenewrobustcmd{\Att}{\mathrel{\NAtt}}
\newrobustcmd{\XAtt}{\withkl{\kl[\Att]}{\NAtt}}
\knowledgenewrobustcmd{\Lab}{\cmdkl{\ell}}

\newcommand{\Scx}{\Phi}
\newcommand{\Scy}{\Psi}
\newcommand{\Scz}{\Xi}
\knowledgenewrobustcmd{\Arg}{\cmdkl{\mathcal{A}}}
\newcommand{\Njust}{\cmdkl{\curvearrowright}}
\knowledgenewrobustcmd{\just}{\mathrel{\Njust}}
\newrobustcmd{\Xjust}{\withkl{\kl[\just]}{\Njust}}
\newcommand{\Ntjust}{\cmdkl{\curvearrowright^*}}
\knowledgenewrobustcmd{\tjust}{\mathrel{\Ntjust}}
\newrobustcmd{\Xtjust}{\withkl{\kl[\tjust]}{\Ntjust}}
\knowledgenewrobustcmd{\self}{{\cmdkl{\circlearrowright}}}
\knowledgenewrobustcmd{\Int}{\cmdkl{\mathcal{I}}}
\newcommand{\Nopen}{\cmdkl{\scalebox{0.8}{$\leftarrow\!\!\rightarrow$}}}
\knowledgenewrobustcmd{\open}{\mathrel{\Nopen}}
\newrobustcmd{\Xopen}{\withkl{\kl[\open]}{\Nopen}}
\newcommand{\Nclos}{\cmdkl{\scalebox{0.8}{$\rightarrow\!\!\leftarrow$}}}
\knowledgenewrobustcmd{\clos}{\mathrel{\Nclos}}
\newrobustcmd{\Xclos}{\withkl{\kl[\clos]}{\Nclos}}
\knowledgenewrobustcmd{\openclos}{\mathrel{\cmdkl{\scalebox{0.8}{$\leftrightarrow\!\!\leftrightarrow$}}}}
\newcommand{\Snx}{\mathfrak{S}}
\newcommand{\Sny}{\mathfrak{T}}
\newcommand{\Snz}{\mathfrak{U}}
\newcommand{\rsf}[1]{\text{\scriptsize$\mathsf{#1}$}}

\newcommand{\limp}{\Rightarrow}

\renewcommand{\emptyset}{\varnothing}

\DeclareRobustCommand{\sys}[1]{$\mathsf{#1}$}

\newcommand{\compr}[2]{\left\{#1 ~\middle|~ #2\right\}}

\knowledgenewrobustcmd{\prnt}[1]{{\cmdkl{\rightrightarrows}}#1} % Parents
\knowledgenewrobustcmd{\chld}[1]{#1{\cmdkl{\rightrightarrows}}} % Children
\newcommand{\Nsibl}{\cmdkl{\bowtie}}
\knowledgenewrobustcmd{\sibl}{\mathrel{\Nsibl}} % Sibling
\newrobustcmd{\Xsibl}{\withkl{\kl[\sibl]}{\Nsibl}}

\knowledgenewrobustcmd{\rsubg}[1]{#1{\cmdkl{\downarrow}}}

\knowledgenewrobustcmd{\prune}[2]{\cmdkl{\mathsf{prune}}(#1, #2)} % Pruning
\knowledgenewrobustcmd{\collapse}[2]{\cmdkl{\mathsf{collapse}}(#1, #2)} % Collapsing

\knowledgenewrobustcmd{\InsNodes}[1]{\cmdkl{\mathsf{Intro}}(#1)}
\knowledgenewrobustcmd{\DelNodes}[1]{\cmdkl{\mathsf{Elim}}(#1)}
\knowledgenewrobustcmd{\OpnNodes}[1]{\cmdkl{\mathsf{Opn}}(#1)}
\knowledgenewrobustcmd{\CloNodes}[1]{\cmdkl{\mathsf{Clo}}(#1)}

\knowledgenewrobustcmd{\dunion}{\mathop{\cmdkl{\uplus}}}
\knowledgenewrobustcmd{\lcpy}[1]{#1^\triangleleft}
\knowledgenewrobustcmd{\rcpy}[1]{#1^\triangleright}

\knowledgenewrobustcmd{\sssiso}{\mathrel{\cmdkl{\simeq}}}

\newcommand{\lequiv}{\simeq}

\knowledgenewrobustcmd\SA{\cmdkl{\textsc{\small SA}}}

\newcommand{\lquote}{\mathopen{\textnormal{\raisebox{0pt}[0pt][0pt]{‹}}}}
\newcommand{\rquote}{\mathclose{\textnormal{\raisebox{0pt}[0pt][0pt]{›}}}}

\knowledgenewrobustcmd{\scroll}[2]{\left[#1\left(#2\right)\right]}
\knowledgenewrobustcmd{\oscroll}[2]{\left[#1\lquote\left(#2\right)\right]}
\knowledgenewrobustcmd{\cscroll}[2]{\left[#1\left(#2\right)\rquote\right]}
\knowledgenewrobustcmd{\ocscroll}[2]{\left[#1\lquote\left(#2\right)\rquote\right]}
\knowledgenewrobustcmd{\vscroll}[2]{\left[#1\left\langle#2\right\rangle\right]}
\knowledgenewrobustcmd{\ovscroll}[2]{\left[#1\left\langle#2\right)\right]}
\knowledgenewrobustcmd{\cvscroll}[2]{\left[#1\left(#2\right\rangle\right]}
\knowledgenewrobustcmd{\ovcscroll}[2]{\left[#1\left\langle#2\right)\rquote\right]}
\knowledgenewrobustcmd{\cvoscroll}[2]{\left[#1\lquote\left(#2\right\rangle\right]}

\newcommand{\SepNode}{\mathbin{\tikz[baseline=-0.5ex]{\fill (0,0) circle (0.65mm);}}}
\tikzset{
  attach/.style={
    ultra thick,
    orange
  }
}

\knowledgenewrobustcmd{\fgtI}[1]{\cmdkl{\downharpoonleft}#1\cmdkl{\downharpoonright}} % Nets -> Interfaces
\knowledgenewrobustcmd{\fgtS}[1]{\cmdkl{\downharpoonleft}#1\cmdkl{\downharpoonright}} % Interfaces -> Structures

\knowledgenewrobustcmd{\src}{\cmdkl{\mathsf{src}}}
\knowledgenewrobustcmd{\tgt}{\cmdkl{\mathsf{tgt}}}

\knowledgenewrobustcmd{\Atoms}{\cmdkl{\mathcalboondox{A}}}
\knowledgenewrobustcmd{\Vars}{\cmdkl{\mathcalboondox{V}}}

\knowledgenewrobustcmd{\jus}[2]{#1\cmdkl{/}#2}
\knowledgenewrobustcmd{\var}[2]{#1\mathop{\cmdkl{:}}#2}
\newcommand{\bdot}{\scalebox{0.4}{$\bullet$}}
\knowledgenewrobustcmd{\bnd}[1]{\overset{\cmdkl{\bdot}}{#1}}

\knowledgenewrobustcmd{\bs}[1]{\mathop{\cmdkl{\raisebox{0.6mm}{\bdot}}} #1}

\newcommand{\dom}{\operatorname{dom}}
\knowledgenewrobustcmd{\vars}{\cmdkl{\mathsf{vars}}}
\knowledgenewrobustcmd{\prem}[1]{\left\lceil #1 \right\rceil}
\knowledgenewrobustcmd{\conc}[1]{\left\lfloor #1 \right\rfloor}

\knowledgenewrobustcmd{\pderiv}[3]{#1 \mathrel{\cmdkl{\vdash}} #2 \mathrel{\cmdkl{\,\vartriangleright\,}} #3}
\knowledgenewrobustcmd{\nderiv}[3]{#1 \mathrel{\cmdkl{\vdash}} #2 \mathrel{\cmdkl{\,\blacktriangleright\,}} #3}
\knowledgenewrobustcmd{\deriv}[2]{#1 \mathrel{\cmdkl{\vartriangleright}} #2}
\newrobustcmd{\Xderiv}{\withkl{\kl[\deriv]}{\cmdkl{\vartriangleright}}}
\knowledgenewrobustcmd{\tderiv}[2]{#1 \mathrel{\cmdkl{\vartriangleright^*}} #2}

\newcommand{\xstep}[1]{\xrightarrow{#1}}

\knowledgenewrobustcmd{\red}[1]{\cmdkl{\rightsquigarrow}_{\mathsf{#1}}}

\knowledgenewrobustcmd{\interp}[1]{\cmdkl{\llbracket}#1\cmdkl{\rrbracket}}
\knowledgenewrobustcmd{\pinterp}[1]{\cmdkl{\llceil}#1\cmdkl{\rrceil}}
\knowledgenewrobustcmd{\cinterp}[1]{\cmdkl{\llfloor}#1\cmdkl{\rrfloor}}

\knowledgenewrobustcmd{\compat}{\mathrel{\cmdkl{\sim}}}

\knowledgenewrobustcmd{\superp}{\mathop{\cmdkl{\gtrdot}}}

\knowledgenewrobustcmd\ltrad[1]{#1^{\cmdkl{\circ}}}

\begin{document}

\maketitle

\begin{abstract}
  We introduce a new formalism for representing proofs in propositional logic called ``scroll
  nets''. Its fundamental construct is the scroll, a topological notation for implication proposed
  by C. S. Peirce at the end of the 19th century as the basis for his diagrammatic system of
  existential graphs (EGs). Scroll nets are derived from EGs by following the Curry-Howard
  methodology of internalizing inference rules inside judgments, just as terms in type theory
  internalize natural deduction rules. We focus on the intuitionistic implicative fragment of EGs,
  starting from a natural diagrammatic representation of scroll nets, and then distilling their
  combinatorial essence into a purely graph-theoretic definition. We also identify a notion of
  detour, that we use to sketch a detour-elimination procedure akin to cut-elimination. We
  illustrate how to simulate normalization in the simply typed $\lambda$-calculus, demonstrating
  both the logical and computational expressivity of our framework.
\end{abstract}

\section{Introduction}

\subparagraph*{Graphical proof theory}
Traditionally, proof theory has been established as a branch of ""symbolic"" logic: it embodies
\textit{par excellence} the Hilbertian ideal according to which any mathematical concept can be
expressed as a finite sequence of "symbols", with reasoning reduced to the sequential manipulation of
these "symbols". However, several recent developments tend to demonstrate that a finer analysis of the
structure of formal proofs is possible with the help of ""graphical"" or diagrammatic
representations: popular examples include Girard's proof nets in linear logic
\cite{girard-linear-1987}, string diagrams in categorical logic \cite{10.1007/11874683_1}, and
Hughes' combinatorial proofs \cite{Hughes_2006}. The idea is to abstract from the arbitrary
sequentiality suggested by a linguistic usage of "symbols", by immersing statements and proofs in
space to better understand their geometry. One aim is to find a solution to Hilbert's 24th problem
by devising a good notion of \emph{equality} among proofs \cite{strasburger-problem-2019}. However,
current research focuses almost exclusively on the structure of \emph{completed} proofs, forgetting
the sequential inference process that enabled their construction. This prevents in particular the
application of these new methodologies to the design of ""interactive theorem provers""
(\reintro{ITPs}), which are fundamentally based on the incremental construction of \emph{partial}
proofs.
\vspace{-0.5em}

\begin{figure}
  \captionsetup[subfigure]{justification=centering}
  \centering
  \begin{subfigure}[b]{0.74\textwidth}
    \scalebox{0.97}{
    \begin{mathpar}
      \includepdf{2cm}{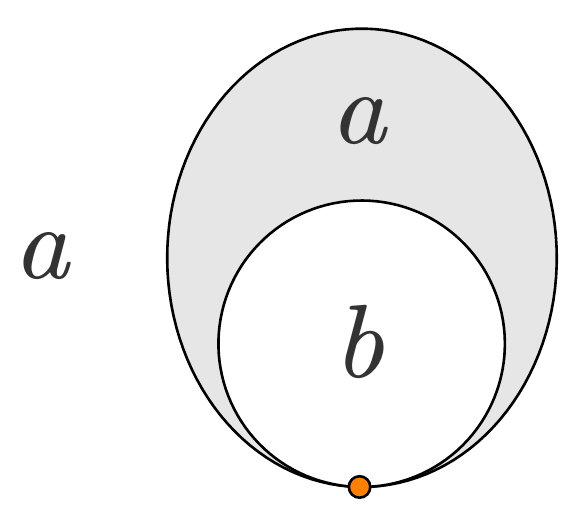}
      ~~~\xstep{"Deit"}~~~
      \includepdf{2cm}{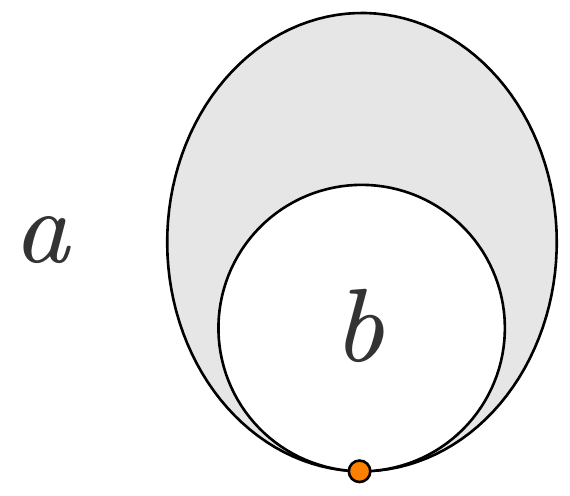}
      ~~~\xstep{"Close"}~~~
      \includepdf{1.25cm}{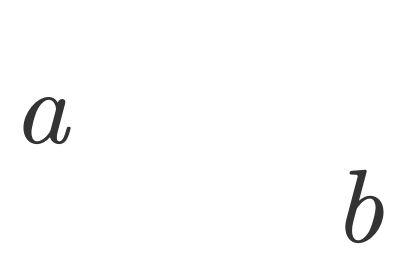}
      ~~~\xstep{"Del"}~~~
      \includepdf{0.25cm}{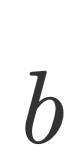}
    \end{mathpar}
    }
    \caption{Dynamic "proof trace"}
    \label{fig:modus-ponens-dynamic}
  \end{subfigure}
  \begin{subfigure}[b]{0.25\textwidth}
    \centering
    $\includepdf{2cm}{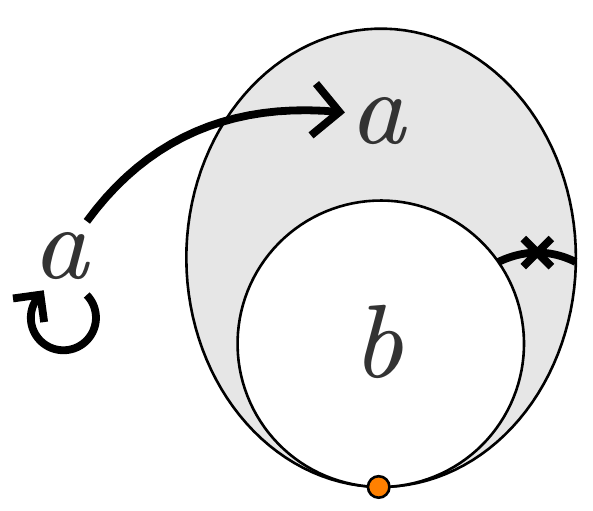}$
    \caption{Static "proof object"}
    \label{fig:modus-ponens-static}
  \end{subfigure}
  \caption{Modus ponens in "scroll nets"}
  \label{fig:modus-ponens}
\end{figure}

\subparagraph*{Existential graphs}
To tackle this limitation, we investigate a "graphical" system mostly forgotten by contemporary proof
theorists, probably because it predates the existence of proof theory itself: C. S. Peirce's
""existential graphs"" \cite{Roberts+1973} (``\reintro{EGs}'' hereafter). It is based on a purely
diagrammatic and topological representation of logical constants, inspired by a dialogical
understanding of reasoning a century before the advent of game semantics
\cite{pietarinenPeircesGametheoreticIdeas2003}. Proving is then modelled as a dynamic process of
constructing valid statements through rewriting of diagrams, using six elementary inference rules
that perform insertions and deletions of "graphs" at a single specified location.
% For reasons that will become clear later on, we call this latter property \emph{illative atomicity}.
Figure \ref{fig:modus-ponens-dynamic} illustrates how to derive \textit{modus ponens} in "EGs",
starting from a "graph" of the statement $a \land (a \limp b)$ and reducing it to $b$. We refer the
reader to \cite[Sections~2--4]{flower-calculus} for a more detailed overview of the history of
"EGs".

\subparagraph*{What is a proof?}
In order to contemplate the structure of (partial) proofs, one needs a way to represent them
\emph{statically} as bona fide ""proof objects"". From our perspective, formalisms like
Hilbert-style and Gentzen-style calculi conflate the final "proof object" with what we call the
""proof trace"" (usually called ``derivation''), which is the history of inference steps that
were undertaken to construct \emph{dynamically} the "proof object". On the other hand, "graphical"
formalisms such as proof nets and combinatorial proofs forego the notion of inference rule
altogether, and with it the whole "proof trace". Although they are also "graphical" in nature, "EGs"
somehow lay on the opposite end of the spectrum, together with string diagrams: they lack explicit
syntax to capture the structure of proofs, leaving it implicit and mangled in the sequence of
rewritings that represents the "proof trace". In contrast, state-of-the-art "ITPs" such as Rocq
\cite{the_rocq_development_team_2025_15149629} and Lean \cite{10.1007/978-3-030-79876-5_37} do have
distinct notions of "proof trace" and "proof object": they are called respectively ""proof script""
and ""proof term"". The full situation is summarized in Table \ref{tab:proof-repr}.

\begin{table}[]
    \centering
    \scalebox{0.87}{
    \def\arraystretch{1.25}%
    \begin{tabular}{l|c|c|ll}
        \cline{2-3}
        \multicolumn{1}{c|}{}                             & \textbf{"Proof trace"}              & \textbf{"Proof object"}                & \multicolumn{1}{c}{\textbf{}} & \multicolumn{1}{c}{\textbf{}} \\ \cline{1-3}
        \multicolumn{1}{|l|}{\textbf{Hilbert calculi}}    & Derivation \emph{(sequence)}      & Derivation \emph{(sequence)}         &                               &                               \\ \cline{1-3}
        \multicolumn{1}{|l|}{\textbf{Gentzen calculi}}    & Derivation \emph{(tree)}          & Derivation \emph{(tree)}             &                               &                               \\ \cline{1-3}
        \multicolumn{1}{|l|}{\textbf{Rocq/Lean}}          & "Proof script" (= tactic \emph{tree})        & "Proof term" (= natural deduction \emph{tree})             &                               &                               \\ \cline{1-3}
        \multicolumn{1}{|l|}{\textbf{Proof nets}}         & ---                               & Formula \emph{tree} + Axiom/Cut \emph{permutation} &                               &                               \\ \cline{1-3}
        \multicolumn{1}{|l|}{\textbf{String diagrams}}    & Equational rewriting (\emph{sequence})     & ---                                  &                               &                               \\ \cline{1-3}
        \multicolumn{1}{|l|}{\textbf{"Existential graphs"}} & "Illative transformations" (\emph{sequence}) & ---                                  &                               &                               \\ \cline{1-3}
        \multicolumn{1}{|l|}{\textbf{"Scroll nets"}}        & "Illative transformations" (\emph{sequence}) & "EG" \emph{"DAG"} + "Argumentation" \emph{forest}        &                               &                               \\ \cline{1-3}
    \end{tabular}
    }
    \caption{Comparing representations of proofs in various formalisms}
    \label{tab:proof-repr}
\end{table}

\subparagraph*{Scroll nets}
The aim of the current paper is to introduce a language of \emph{proofs} on top of the existing
language of statements in "EGs", in order to represent "proof objects" in addition to "proof traces".  A
notable feature of "EGs" is that they identify the language of statements with the language of
\emph{judgments}: scribing some "graph" on the \emph{"sheet of assertion"} is, by definition, the same
as asserting its truth\footnote{This identification is also found in the original notation of
Gentzen for natural deduction, where nodes in derivation trees are formulas rather than sequents
\cite{girardProofsTypes1989}.}. Following the Curry-Howard methodology, our goal is then to
internalize inference rules --- Peirce called them ""illative transformations"" --- inside the
syntax of statements, just like terms in type theory can be seen as a way to internalize and record
derivation trees (in natural deduction) inside judgments.

% A consequence of the judgment--statement identification is that the rich structure of statements
% should be reflected in the structure of illative transformations, and in their static representation
% thereof. This will make our syntax closer to \emph{deep inference} formalisms like the calculus of
% structures \cite{Guglielmi1999ACO} and open deduction \cite{deep_inference}\footnote{And also
% categorical proof theory as noted in \cite{hughesDeepInferenceProof}.}, where proofs can also be
% composed by connectives. Hence a traditional term syntax is likely to fail at faithfully capturing
% this rich structure, since the tree shape of terms is in exact correspondence with the shallow
% structure of usual inference rules acting on sequents\footnote{Unfortunately this is the research
% path that was followed by the author for almost 2 years...}.

The syntax we have arrived at is based on the simple observation that all "illative transformations"
can be expressed in terms of pure \emph{locations}: they consist in either \emph{inserting/deleting}
an arbitrary "graph" at a given location, \emph{duplicating} an arbitrary "graph" from a source
location, or \emph{deduplicating} an arbitrary "graph" from a target location. This gives rise to a
directed forest whose nodes are the ``"subgraphs"'' of some "EG" and whose edges encode exactly
"illative transformations". This data structure is surprisingly close to the \emph{proof nets} of
linear logic, and because it is based on the fundamental construct of the \emph{"scroll"} coming
from intuitionistic "EGs", we call it \textbf{""scroll net""}. Figure~\ref{fig:modus-ponens-static}
shows an example of "scroll net", which is exactly the "proof object" built by the derivation of
Figure~\ref{fig:modus-ponens-dynamic}.
   
% \paragraph{Ecumenism}
% More than a mere aesthetic principle, our current (ongoing) work suggests that the aforementioned
% duality provides a very general setting to capture not only provability in classical or
% intuitionistic logic, but also the fine structure of proofs in many flavors of intermediate and
% substructural logics. Importantly, our emerging framework forces all these logics to share the same
% language of statements (that of existential graphs), identifying each logic as a subset of proofs
% with specific topological and computational properties. This should allow for a new approach to
% \emph{ecumenism} in logic, differing from previous attempts such as Girard's Logic of Unity
% \cite{logicUnity}, Prawitz' Ecunemical Logic \cite{marinEcumenicalModalLogic2020} or even logical
% frameworks like LF \cite{harperFrameworkDefiningLogics1993} by the abandonment of symbolic
% connectives in favor of a unified set of diagrammatic signs.

\subparagraph*{Outline}
The article is organized as follows: in Section~\ref{sec:IEGs} we recall the diagrammatic syntax of
the implication-conjunction fragment of intuitionistic "EGs", based on the fundamental icon of the
"scroll". In Section~\ref{sec:reif} we explain how to internalize "illative transformations" inside
"EGs" by using a diagrammatic arrow notation. In Section~\ref{sec:combinatorial} we give a formal
combinatorial definition of "scroll nets", based on a "DAG" generalization of "EGs" combined with a
forest of "illative transformations".
% In Section~\ref{sec:inductive} we provide an inductive syntax and textual notation based on this
% characterization, that we use to define horizontal and vertical composition operations on "scroll
% nets".
In Section~\ref{sec:soundness} we give a sequential "correctness" criterion for "scroll nets"
following the dynamic understanding of "illative transformations", that we use to define
"horizontal" and "vertical composition" operations.
% , based on an inductive specification of derivability in the style of small-step operational
% semantics that is shown to preserve logical soundness.
In Section~\ref{sec:computation} we identify four kinds of "detours" that arise when a node is both
"introduced" and "eliminated", and sketch informally a "detour" elimination procedure. We also give
a direct translation of simply typed $\lambda$-calculus into "scroll nets", illustrating how to
simulate $\beta$-reduction. We conclude in Section~\ref{sec:conclusion} with a discussion of related
works and future directions to improve the metatheory of our framework, extend it to richer logics,
and use it in "interactive theorem proving" applications.

\section{Implicative existential graphs}\label{sec:IEGs}

\subparagraph*{Sheet of Assertion}
The most fundamental concept of "EGs" is the \intro{sheet of assertion}, denoted by $\intro*\SA$
thereafter. It is the space where statements are scribed by the reasoner, typically a sheet of
paper, a blackboard, or a computer display. This last analogy suggests an important property of
$\SA$: it must offer a \emph{virtually infinite} amount of space, so that one can perform as much
reasoning as needed by scribing an unbounded (but finite) amount of statements\footnote{Just like a
Turing machine has an infinite tape, so that one can perform as much computation as needed. In
\kl{symbolic} logic, this is captured by the fact that formulas, although usually finite, can have
an unbounded size.}.

Then as the name indicates, scribing some statement $\Scx$ on $\SA$ has the meaning of
\emph{asserting the truth of} $\Scx$. It is thus an instance of the notion of \emph{judgment} as
identified by logicians like Frege and Martin-Löf, who would write it "symbolically" as $\vdash \Scx$.
Naturally, the empty $\SA$ in interpreted as the absence of any assertion/judgment, and thus as
vacuous truth $\top$.

\subparagraph*{Atoms}
Peirce often used sentences expressed in natural language as the most elementary statements scribed
on $\SA$, probably for pedagogical purposes. However he made clear that the informal meaning of
these sentences, that is their denotation in the real world, is irrelevant to the process of pure
logical deduction. This agrees with the modern view on \emph{atomic} propositions, which are taken
to be arbitrary abstract "symbols" drawn from some countably infinite set $\intro*\Atoms$. We will
use letters $a, b, c\ldots$ to denote such statements and call them \intro{atoms}.

\subparagraph*{Juxtaposition}
Recall that one can scribe an arbitrary number of statements on $\SA$, thus asserting the truth of
each of them simultaneously. That is, \intro{juxtaposition} has the meaning of \emph{conjunction}, as we know from
the introduction rule for $\land$ in natural deduction. However, "symbolic" connectives do not exist
in the syntax of "EGs", because Peirce aimed precisely for a "symbolless" --- what he called
\intro{iconic}~\cite{10.7551/mitpress/3633.001.0001} --- notation for logic. Thus very concisely,
$$
\vcenter{\prftree[r]{$\land\mathsf{i}$}{\vdash \Scx}{\vdash \Scy}{\vdash \Scx \land \Scy}} \qquad\text{is expressed by the "graph"}\qquad \Scx \quad \Scy
$$
Note that in "symbolic" logic, $\Scx$ and $\Scy$ can be arbitrarily complex formulas, not just
"atoms". In "EGs", a complex statement --- that we will call a ""graph"" --- is any delimited
portion/area of $\SA$, as long as the delimitation does not interrupt the continuity of some
""token"".

\subparagraph*{Scroll}
In the fragment of "EGs" considered in this article, a "token"\footnote{Peirce used the word
``"token"'' as a synonym of the more contemporary term ``occurrence'', and the word ``type'' to refer
to the common pattern that is instantiated in multiple occurrences of the same type.} is a scribed
occurrence of either an "atom", or what Peirce called a ""scroll"". In the words of Peirce
himself~\cite[pp.~533--534]{peirce_prolegomena_1906}:
\begin{quote}
  Accordingly, since logic has primarily in view argument, and since the conclusiveness of an
argument can never be weakened by adding to the premisses or by subtracting from the conclusion, I
thought I ought to take the general form of argument as the basal form of composition of signs in my
diagrammatization; and this necessarily took the form of a ``scroll'', that is \textelp{} a curved
line without contrary flexure and returning into itself after once crossing itself, and thus forming
an outer and an inner ``close''.
\end{quote}

Examples of "scrolls" can be found in Figure~\ref{fig:modus-ponens}, where the curved line is depicted
by a white ellipse (the ``inner close'') nested in a gray ellipse (the ``outer close''), with a
unique orange intersection point emphasizing the ``once crossing itself'' part. Peirce also used the
term ""sep"" instead of ``close'' to refer to any of the two closed curves that make up a
"scroll"\footnote{This is because a closed curve literally \emph{"sep"}arates $\SA$ into two distinct
areas, which is now known as the \emph{Jordan curve theorem} in topology.}. Following Pietarinen
\cite{minghui_graphical_2019}, we will use the terms ""inloop"" and ""outloop"" to designate the
inner and outer "seps" of a "scroll", respectively. By ``general form of argument'', Peirce alludes
to logical \emph{implication} $\Scx \limp \Scy$, where the antecedant $\Scx$ resides in the "outloop",
and the consequent $\Scy$ in the "inloop". 
% $$
% \vcenter{\prftree[r]{${\limp}\mathsf{i}$}{\Scx \vdash \Scy}{\vdash \Scx \limp \Scy}} \qquad\text{is expressed by the graph}\qquad \includepdf{2cm}{scroll}
% $$
% Notice that as in the case of conjunction/juxtaposition, occurrences of $\Scx$ and $\Scy$ that are
% morally ``the same'' in the introduction rule become \emph{identified} in "EG" notation. We will
% come back to this remarkable compactness of the representation in Section~\ref{sec:conclusion},
% while the fully general translation of natural deduction rules will be given in Section~\ref{sec:stlc}.
$\Scx$ and $\Scy$ can be arbitrarily complex "graphs", so that in particular a "scroll" located in
the "outloop" of another "scroll" will be drawn with its gray and white shading inverted in order to
reflect the change of polarity (see for instance
Figure~\ref{fig:annotated-interaction-rules}).
% so that $(a \land (a \limp b)) \limp b$ would be depicted by the following "graph":
% $$\includepdf{2.25cm}{scroll}$$
% Since they denote not only implication but also \emph{entailment}, Pietarinen noticed that this can
% be understood as a form of \emph{nested sequents} \cite{minghui_graphical_2019}.

Finally, note that in classical logic "seps" can be interpreted as \emph{negations} because of the
classical equivalence $\Scx \limp \Scy \lequiv \neg(\Scx \land \neg \Scy)$. Indeed "EGs" were
invented around 1896 before the advent of intuitionistic logic, so that Peirce had a boolean
interpretation of logical operations in mind. However in this article, we will demonstrate that the
inference rules of "EGs" are not only intuitionistically sound (when restricted to "scrolls"), but
that the notion of computation that emerges from them is very much in line with the modern
conception of constructivism.

\begin{figure}
  \captionsetup[subfigure]{justification=centering}
  \centering
  \begin{subfigure}[b]{0.29\textwidth}
    \begin{align*} 
      \Scx \,\,&\xstep{"Open"}\,\, \includepdf{1.1cm}{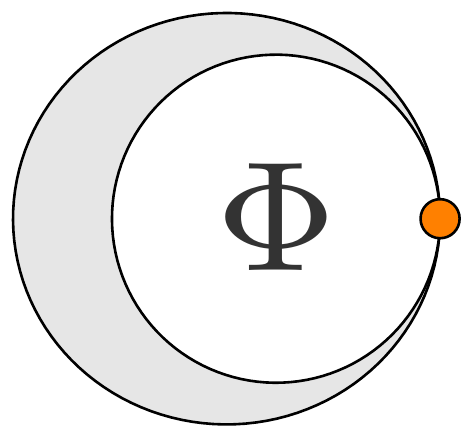} \\
      \includepdf{1.1cm}{opening} \,\,&\xstep{"Close"}\,\, \Scx
    \end{align*}
    \caption{Interaction rules}
    \label{fig:interaction-rules}
  \end{subfigure}
  \begin{subfigure}[b]{0.65\textwidth}
    \begin{align*} 
      \includepdf{1.3cm}{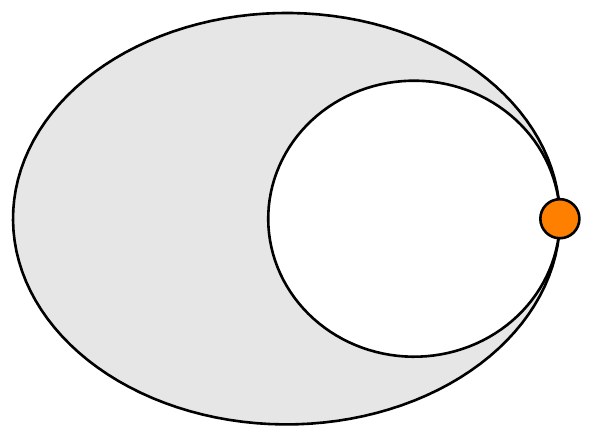} \,\,&\xstep{"Ins"}\,\, \includepdf{1.3cm}{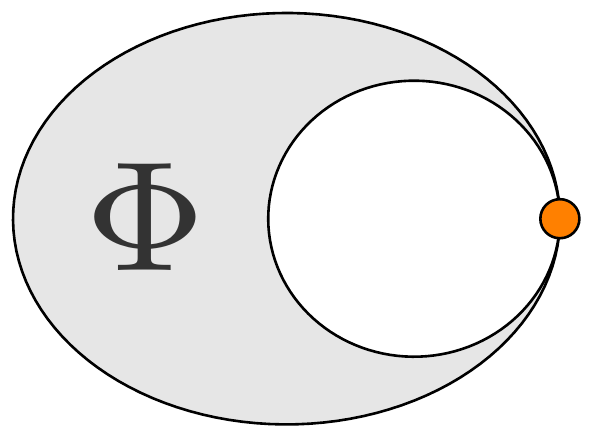} &
      \includepdf{1.3cm}{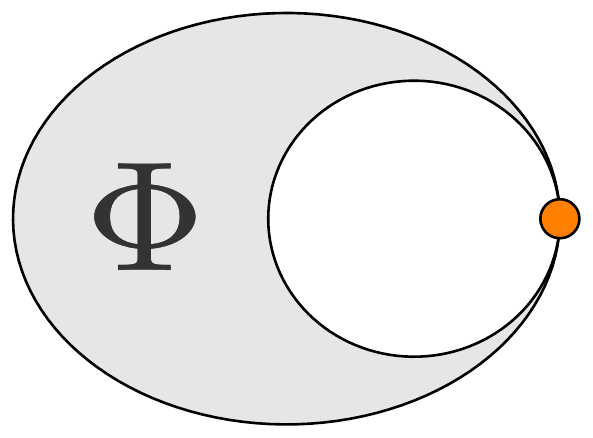} \,\,&\xstep{"Iter"}\,\, \includepdf{1.3cm}{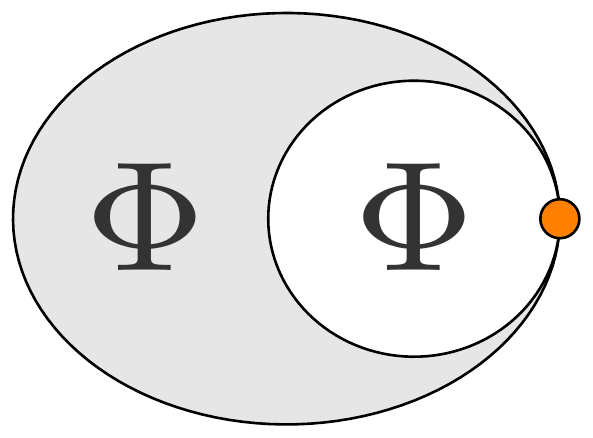} \\
      \includepdf{1.3cm}{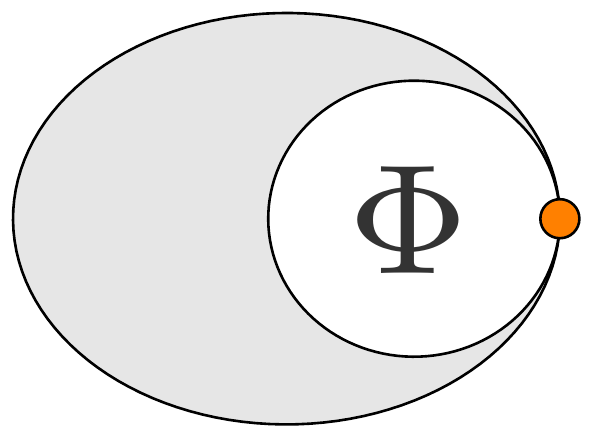} \,\,&\xstep{"Del"}\,\, \includepdf{1.3cm}{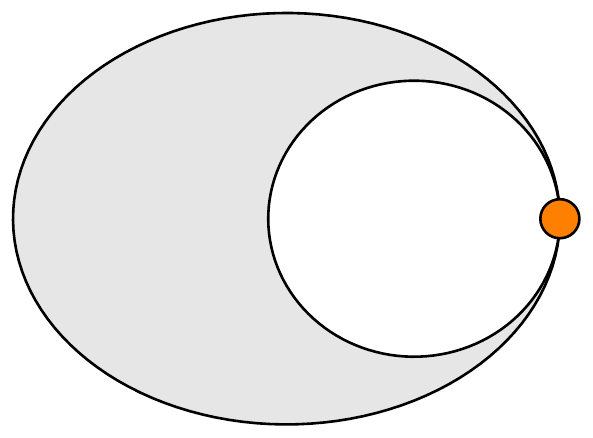} &
      \includepdf{1.5cm}{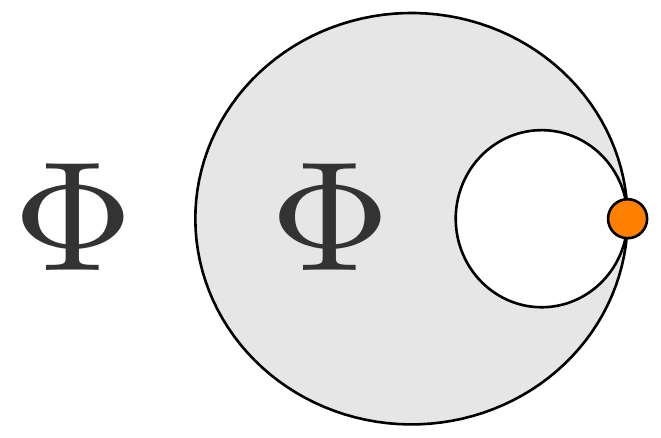} \,\,&\xstep{"Deit"}\,\, \includepdf{1.5cm}{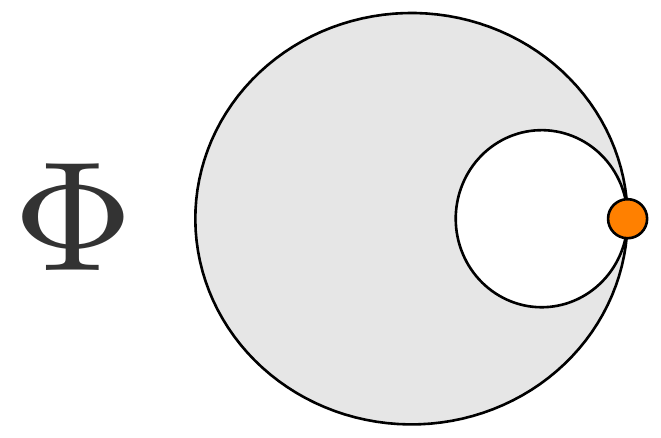}
    \end{align*}
    \caption{Argumentation rules}
    \label{fig:argumentation-rules}
  \end{subfigure}
  \caption{Dynamic "illative transformations"}
  \label{fig:illative-transformations}
\end{figure}

\subparagraph*{Illative transformations}
Peirce called the inference rules of "EGs" \emph{"illative transformations"}. They are traditionally
understood as \emph{rewriting rules}, similarly to the rules of string diagram calculi in category
theory, or to those of the calculus of structures in deep inference \cite{Guglielmi1999ACO}. One can
find various presentations in the literature, some of the rules being \emph{equational} in nature
(bidirectional) because of the equivalence between premiss and conclusion, the others being
necessarily \emph{oriented} (unidirectional). Here we stick to a fully oriented presentation, for
reasons that will become clear in the next section. This makes for a total of six rules, illustrated
with representative examples in Figure~\ref{fig:illative-transformations}. They should be read from
left to right as expressing \emph{forward reasoning} from premiss to conclusion.

While still informal, the following description specifies the rules in their full generality. Note
that to avoid any ambiguity --- and to foster the parallel with natural deduction --- we will use
the terms ""introduction"" and ""elimination"" to refer respectively to the act of \emph{scribing}
and \emph{erasing} a graph from $\SA$. We will also refer to white/gray-shaded areas --- or
alternatively to areas enclosed in an even/odd number of "seps" --- as ""positive""/""negative"".

\begin{description}
  \item[""Opening"" (\intro{Open})]
  A "scroll" with empty "outloop" can be "introduced" around any "graph" $\Scx$.
  \item[""Closing"" (\intro{Close})]
  Any "scroll" with empty "outloop" can be "eliminated".
  \item[""Insertion"" (\intro{Ins})]
  Any "graph" $\Scx$ can be "introduced" in a "negative" location.
  \item[""Deletion"" (\intro{Del})]
  Any "graph" $\Scx$ can be "eliminated" from a "positive" location.
  \item[""Iteration"" (\intro{Iter})] Any "graph" $\Scx$ can be "introduced" in a "positive"
  location, as long as $\Scx$ already occurs in the area of a "sep" that contains said location.
  \item[""Deiteration"" (\intro{Deit})] Any "graph" $\Scx$ can be "eliminated" from a
  "negative" location, as long as $\Scx$ already occurs in the area of a "sep" that contains
  said location.
\end{description}

A remarkable feat of Peirce's rules --- on which he insisted very much --- is that they are only
expressed in terms of "introductions" (first row) and "eliminations" (second row) of "graphs" on
$\SA$. We will refer to this property as ""illative atomicity"". Indeed, Peirce thought that those
were the \emph{smallest} steps in which reasoning could be dissected, making his system extremely
appropriate for \emph{analytical} purposes. This is summarized in the following excerpt
\cite[p.~533]{peirce_prolegomena_1906}:

\begin{quote}
  In the first place, the most perfectly analytical system of representing
propositions must enable us to separate illative transformations into
indecomposable parts. Hence, an illative transformation from any proposition, A,
to any other, B, must in such a system consist in first transforming A into AB,
followed by the transformation of AB into B. For an omission and an insertion
appear to be indecomposable transformations and the only indecomposable
transformations.
\end{quote}

We qualify the first two rules "Open" and "Close" as ""interaction"" rules. Indeed from a
game-semantical point of view, the "Open" rule starts an interaction with the opponent by
introducing a "negative" (resp. "positive") "outloop" in a "positive" (resp. "negative") area, while the
"Close" rule ends this interaction by erasing the "scroll" while keeping the "inloop"'s content.
They are obviously sound in both intuitionistic and classical logic, as they correspond to the
equivalence $\top \limp \Scx \lequiv \Scx$.

We qualify the other rules as ""argumentation"" rules, following Peirce's description of the
"scroll" as the \enquote*{general form of argument}. They generalize exactly the \emph{structural}
rules found in sequent calculus and the calculus of structures (as well as the \emph{switch} rule in
the latter), because they can be applied in locations of arbitrary depth and polarity, not just at
the top-level of sequents or in "positive" contexts.

The "Ins" and "Del" rules correspond respectively to \emph{weakening} and \emph{coweakening}. They
capture the monotonicity of entailment, expressed by Peirce in the sentence \enquote*{\textelp{} the
conclusiveness of an argument can never be weakened by adding to the premisses [("Ins" rule)] or by
subtracting from the conclusion [("Del" rule)] \textelp{}}.

Lastly, the "(de)iteration" rules "Iter" and "Deit" can be seen respectively as generalizations of the
\emph{cocontraction} and \emph{contraction} rules. They capture when some source occurrence of a
statement $\Scx$ \emph{justifies} a distinct target occurrence of the same $\Scx$, where
""justification"" consists in either adding a "positive" conclusion ("Iter") or removing a "negative"
assumption ("Deit"). Their soundness is less straightforward than for other rules, and for lack of space we rely on secondary sources like \cite{flower-calculus} where it has been proved formally with respect to intuitionistic Kripke semantics, albeit in a slightly different setting.

\section{Reifying inference}\label{sec:reif}

\begin{figure}
  \captionsetup[subfigure]{justification=centering}
  \centering
  % \begin{subfigure}[b]{0.25\textwidth}
  \begin{subfigure}[b]{0.49\textwidth}
    $
    \begin{aligned} 
      \includepdf{0.35cm}{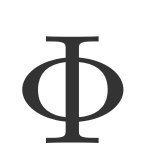} \,&\xstep{}\, \includepdf{1.1cm}{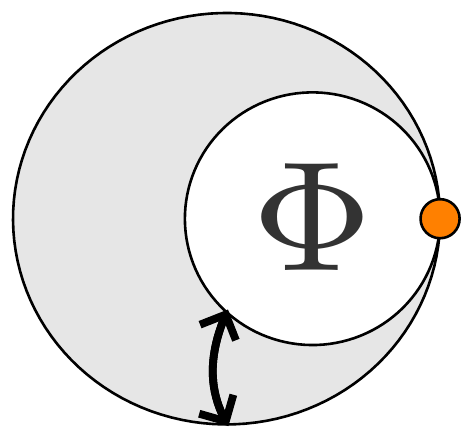} &
      \includepdf{0.9cm}{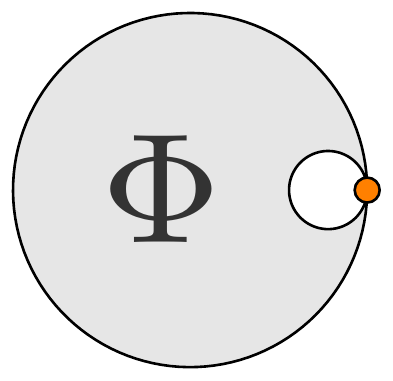} \,&\xstep{}\, \includepdf{1.3cm}{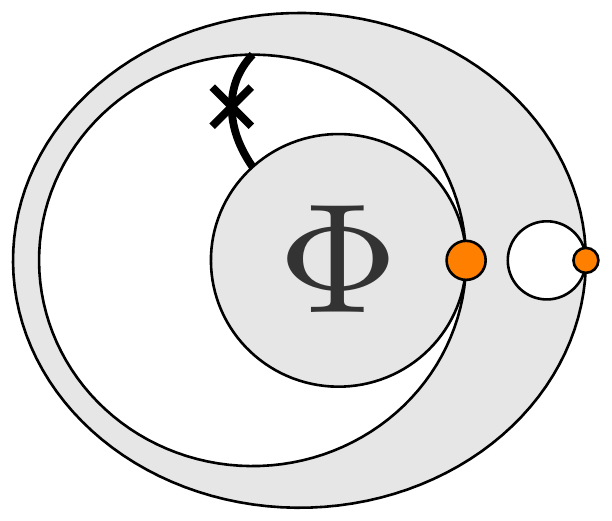} \\
      &\rsf{"Open"+} & &\rsf{"Open"-} \\
      \includepdf{1.3cm}{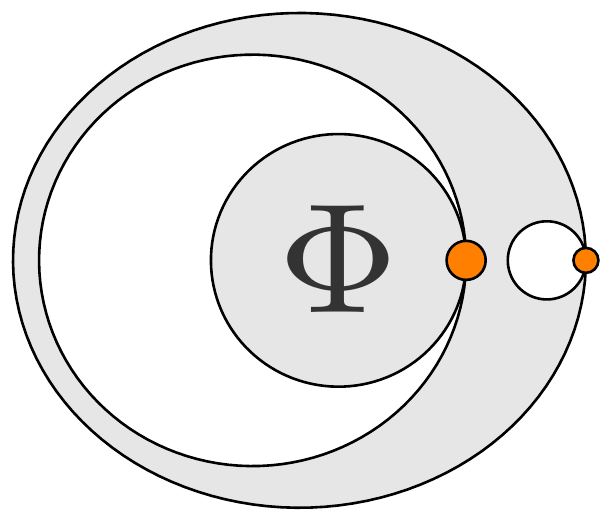} \,&\xstep{}\, \includepdf{1.3cm}{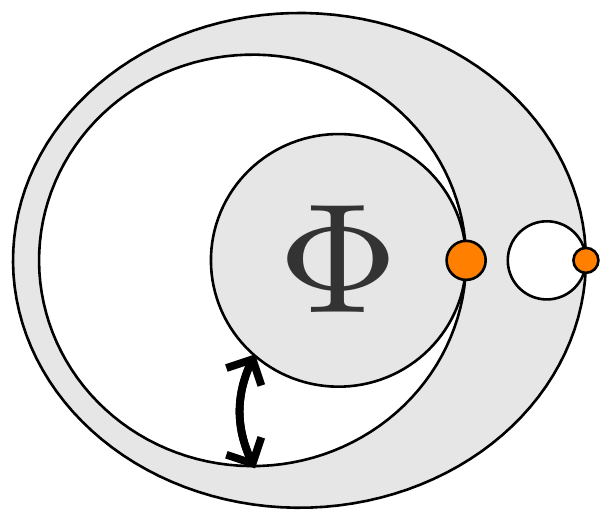} &
      \includepdf{1.1cm}{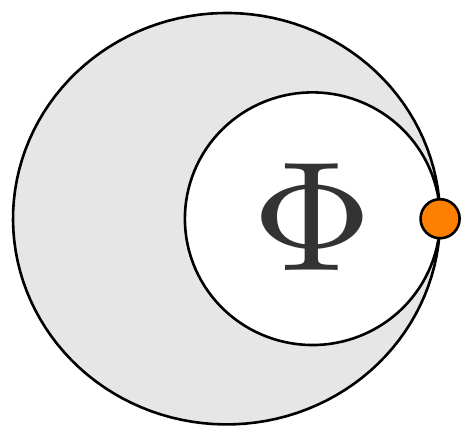} \,&\xstep{}\, \includepdf{1.1cm}{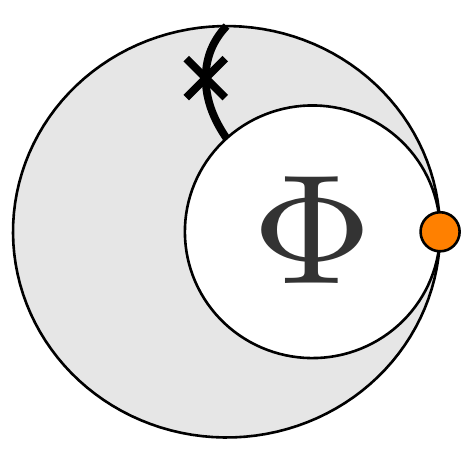} \\
      &\rsf{"Close"-} & &\rsf{"Close"+}
    \end{aligned}
    $
    % $
    % \begin{array}{cc}
    %   \includepdf{1.3cm}{openingp2} & \includepdf{1.5cm}{openingn2} \\
    %   ""Open+"" & ""Open-"" \\
    %   \includepdf{1.3cm}{closingp2} & \includepdf{1.5cm}{closingn2} \\
    %   ""Close+"" & ""Close-""
    % \end{array}
    % $
    \caption{Interaction rules}
    \label{fig:annotated-interaction-rules}
  \end{subfigure}
  % \begin{subfigure}[b]{0.25\textwidth}
  \begin{subfigure}[b]{0.49\textwidth}
    $
    \begin{aligned} 
      \includepdf{1cm}{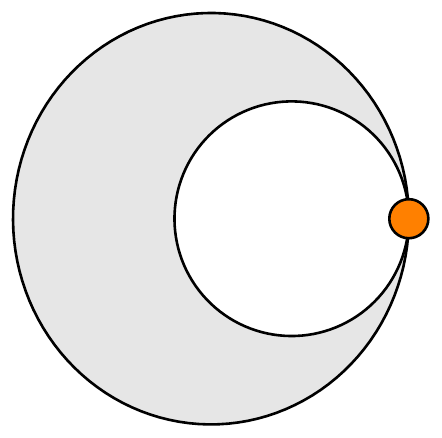} \,&\xstep{}\, \includepdf{1.1cm}{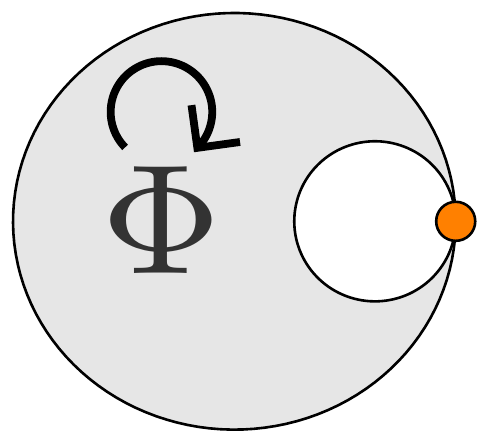} &
      \includepdf{1.3cm}{iteration1} \,&\xstep{}\, \includepdf{1.3cm}{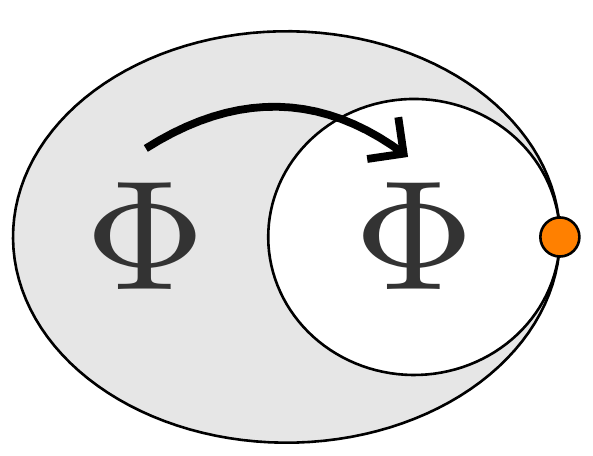} \\
      &\rsf{"Ins"} & &\rsf{"Iter"} \\
      \includepdf{1cm}{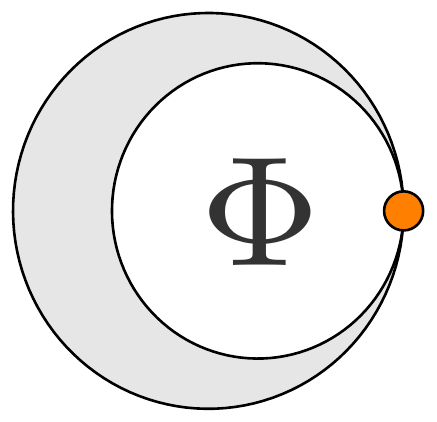} \,&\xstep{}\, \includepdf{1.1cm}{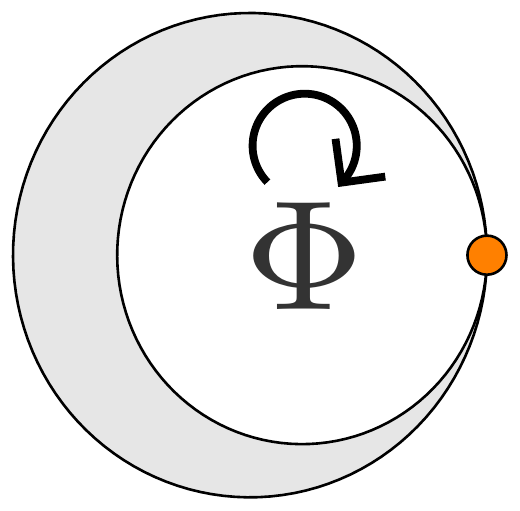} &
      \includepdf{1.2cm}{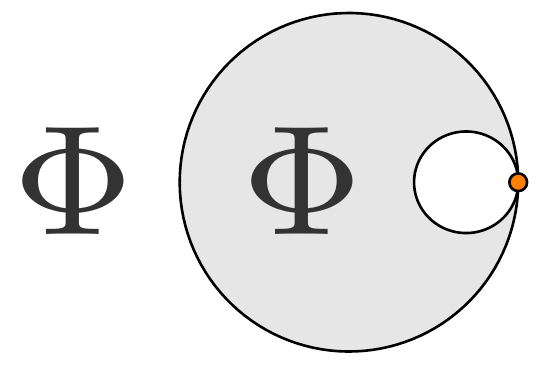} \,&\xstep{}\, \includepdf{1.2cm}{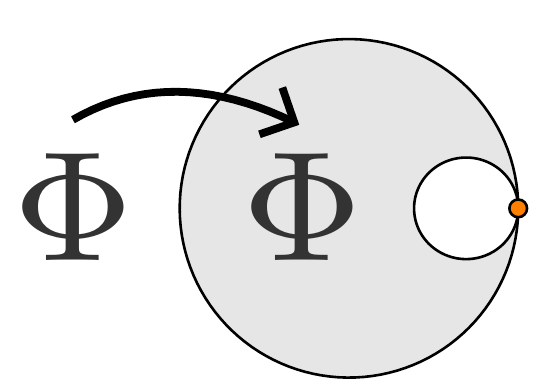} \\
      &\rsf{"Del"} & &\rsf{"Deit"}
    \end{aligned}
    $
    % $
    % \begin{array}{cc}
    %   \includepdf{1.2cm}{insertion2a} & \includepdf{1.5cm}{iteration2a} \\
    %   \rsf{"Ins"} & \rsf{"Iter"} \\
    %   \includepdf{1.3cm}{deletion2a} & \includepdf{1.4cm}{deiteration2a} \\
    %   \rsf{"Del"} & \rsf{"Deit"}
    % \end{array}
    % $
    \caption{Argumentation rules}
    \label{fig:annotated-argumentation-rules}
  \end{subfigure}
  \caption{Static "illative transformations"}
  \label{fig:annotated-illative-transformations}
\end{figure}

\subparagraph*{Justifications} In all presentations of "illative transformations" found in secondary
literature, iteration and deiteration are \emph{polarity-independent}: they can each justify
statements in both "positive" and "negative" locations. Then deiteration can be seen as the converse of
iteration, and since the premiss and conclusion are equivalent, they are usually considered as a
single equational rule. However, Peirce himself noted that this presentation is
\textquote*[{\cite[p.~536]{peirce_prolegomena_1906}}]{\textelp{} valid, sufficient for its purpose,
and convenient in practice, \textelp{but there is a} more scientific way \textins{to proceed}}. Then
he goes on to explain this ``more scientific way'', of which one clause reads as
follows~\cite[p.~537]{peirce_prolegomena_1906}:
\begin{quote}
  \textelp{} if $\Omega$ be a \textbf{recto \textins{(i.e. "positive")}} area, any simple Graph
already scribed upon $Y$ may be iterated upon $\Omega$; while if $\Omega$ be a \textbf{verso
\textins{(i.e. "negative")}} Area, any simple Graph already scribed upon $Y$ and iterated upon
$\Omega$ may be deiterated by being deleted or abolished from $\Omega$.
\end{quote}
This subtle polarity restriction turns out to be very important in order to \emph{reify}
"(de)iteration" in the syntax of "EGs". Indeed, recall that our goal is to find a way to \emph{record}
"illative transformations" on $\SA$ as they are being performed, so that the result of a sequence of
such transformations is the static "proof object" that was built, rather than its mere conclusion;
just as the conclusion of a derivation in type theory is a judgment that holds not only a type, but
also a term witnessing this type.

Then, since a "(de)iteration" justifies a target occurrence of some "graph" $\Scx$ with a source
occurrence of $\Scx$, it is tempting to represent it by a simple \emph{arrow} with corresponding
source and target. This gives the \emph{static} "Iter" and "Deit" rules of
Figure~\ref{fig:annotated-argumentation-rules}. Note that since we want all transformations to
strictly \emph{add} information incrementally, the justified occurrence of $\Scx$ in "Deit" is not
"eliminated" anymore. Then the only way to distinguish between these static versions of "Iter" and
"Deit" is to look at the polarity of the target location.

The same principle can be applied to the "Ins" and "Del" rules, which are now represented by
\emph{looping arrows}. The intuition is that a "graph" "introduced" in a "negative" area with the
"Ins" rule corresponds to an \emph{assumption}, which is ""self-justified"" in the sense that the
reasoner decides it does not require further justification (avoiding the well-known infinite
regression problem). Dually, a "graph" "eliminated" from a "positive" area with the "Del" rule
\emph{annihilates itself}, capturing the reasoner's intent to prevent further usage of the knowledge
about this "graph"'s truth.

\subparagraph*{Interactions}
There is a certain \emph{homeomorphic} flavor to "interaction" rules, expressed by Peirce for the
"Close" rule as follows: \textquote*[{\cite[p.~534]{peirce_prolegomena_1906}}]{\textelp{} the two
walls \textins{(i.e. "seps")} of the scroll, when nothing is between them, fall together,
\textbf{collapse}, disappear, and leave only the contents of the inner close standing \textelp{}}.
We diagrammatize this ``collapse'' deformation by drawing two arrows departing from each "sep" and
meeting at their tips, which also gives the impression of a \emph{cross} symbolizing "elimination"
(rule $\rsf{"Close"+}$ in Figure~\ref{fig:annotated-interaction-rules}). Dually, "opening" (rule
\rsf{"Open"+}) is symbolized with a double-ended arrow whose tips touch the two "seps", evoking an
\emph{expansion} movement as if $\SA$ was teared apart to create "negative" space. In order to stay
consistent with the polarity-dependent interpretation of arrows in "argumentation" rules, collapse
(resp. expansion) arrows are interpreted as "opening" (resp. "closing") transformations in "negative"
areas, giving two new variants \rsf{"Open"-} and \rsf{"Close"-} of the rules.

\section{Combinatorial scroll nets}\label{sec:combinatorial}

\begin{figure}
  \captionsetup[subfigure]{justification=centering}
  \centering
  \begin{subfigure}[b]{0.40\textwidth}
    $$\includegraphics[width=\textwidth]{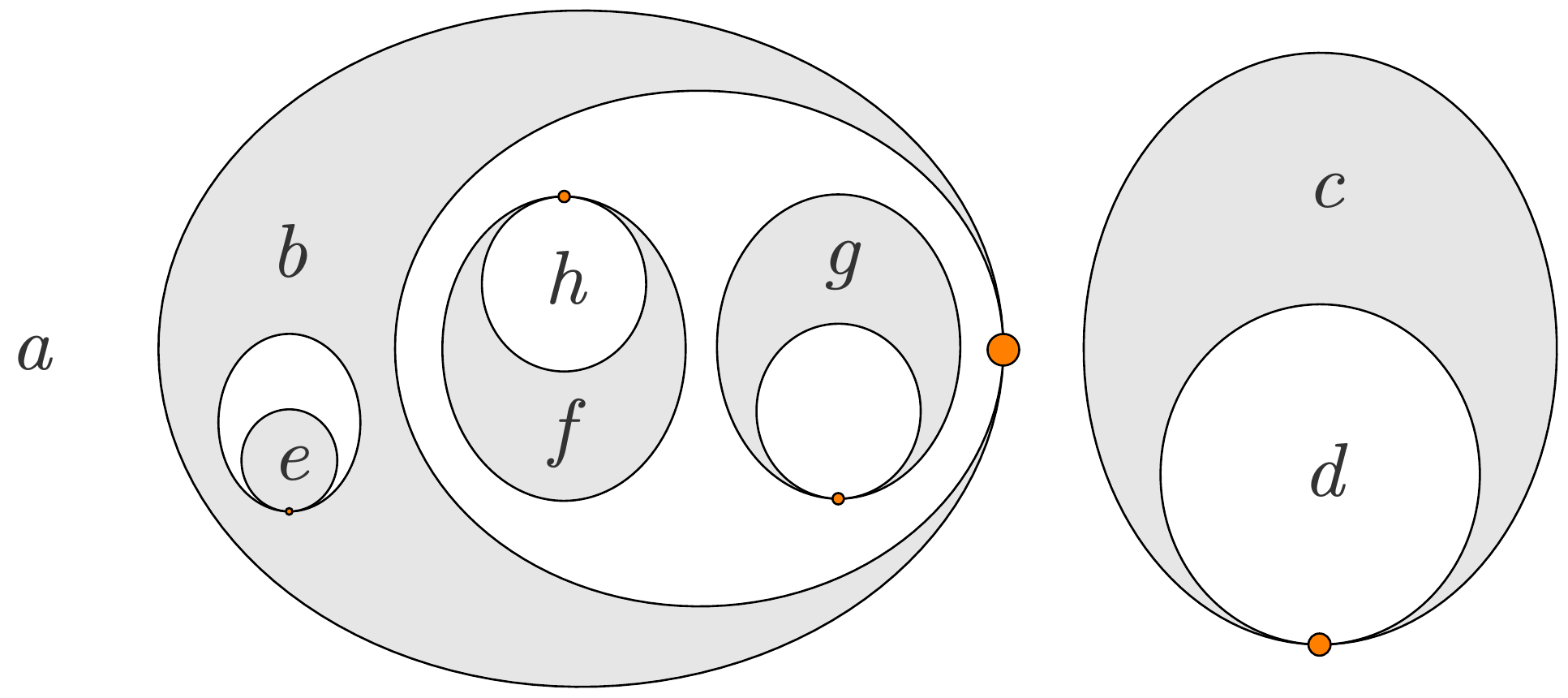}$$
    \caption{Topological}
  \end{subfigure}
  \hspace{2em}
  \begin{subfigure}[b]{0.40\textwidth}
    $$
    \stkfig{0.85}{combinatorial}
    $$
    \caption{Combinatorial}
    \label{fig:combinatorial}
  \end{subfigure}
  \caption{Topological and combinatorial representations of a "simple scroll structure". In
  Fig.~\ref{fig:combinatorial}, labelled leaves are "atoms", $\SepNode$-nodes are "seps", and
  orange edges are "attachments" of "inloops" to "outloops".}
  \label{fig:simple-scroll-structure}
\end{figure}

\subparagraph*{Scroll structures}
As noted by various authors\footnote{See for instance the Tree Existential Graphs of Roberts and
Pronovost \cite{roberts_existential_1992}, or \cite[Section~2.2]{brady_categorical_2000}.}, the
nesting of "seps" on $\SA$ induces a \emph{forest} structure on "EGs": each "sep" constitutes a
node, whose children are either leaves corresponding to "atoms" or empty "seps" residing in the area
of the "sep", or nodes corresponding to nested non-empty "seps". One also needs a way to keep track
of the attachment of "inloops" to "outloops": Figure~\ref{fig:combinatorial} illustrates how this
can be achieved by coloring edges that relate an "inloop" to its parent "outloop" in orange. This
gives the following formal definition:

\begin{definition}[Scroll structure]\label{def:scroll-structure}
A ""simple scroll structure"" is a triple $\Scx = \langle \Gra, \Lab, \XAtt \rangle$ where:
\begin{itemize}
    \item $\intro*\Gra = \langle \intro*\Ver, \intro*\XEdg \rangle$ is a finite rooted
    out-forest (directed from roots to leaves) with edge set $\XEdg \subseteq \Ver \times
    \Ver$. We write $v \Edg u$ for $(v,u) \in \XEdg$, and $\intro*\XtEdg{}$ for the
    transitive closure of ${\XEdg}$.
    \item $\intro*\Lab: L \rightharpoonup \Atoms$ is a partial function labelling the leaves $L \subseteq
    V$ with "atoms".
    \item ${\intro*\XAtt} \subseteq {\XEdg}$ is a subset of ""attachments"" satisfying the following
    well-formedness conditions:
    \begin{enumerate}
        \item ("Atoms" are not "inloops") If $u \Att v$ then $v \notin \dom(\ell)$.
        \item (Every "sep" is attached) $\forall v \in \Ver \setminus \operatorname{dom}(\ell)$,
        $\exists! u \in \Ver$ such that either $u \Att v$ or $v \Att u$.
    \end{enumerate}
\end{itemize}
\end{definition}

We will use letters $\Scx, \Scy, \Scz$ to range over "simple scroll structures", subscripting
components accordingly for disambiguation (e.g. $\Ver_\Scx$ for the vertices of $\Scx$). The adjective
``simple'' is here meant as a reference to \emph{simple type theory}, as will become clear in
Section~\ref{sec:computation}; we will usually omit it for conciseness.

\begin{definition}[Polarity]
  We say that a node $v \in \Ver_\Scx$ is \reintro{positive} (resp. \reintro{negative}) if its
  distance from a root is even (resp. odd). We denote the sets of "positive" and "negative" nodes of
  $\Scx$ by $\intro*\pVer{\Scx}$ and $\intro*\nVer{\Scx}$, respectively.
\end{definition}

\begin{figure}
  \captionsetup[subfigure]{justification=centering}
  \centering
  \begin{subfigure}[b]{\textwidth}
    \centering
    $
      \includepdf{2.35cm}{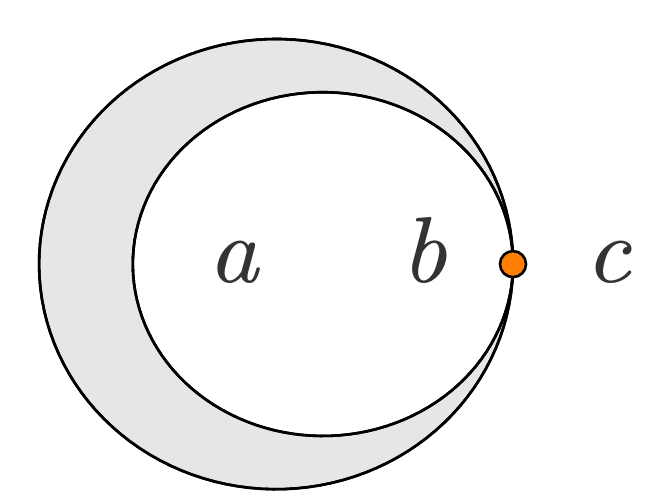}
      ~~~\xstep{"Close"}~~~
      \raisebox{1mm}{$\includepdf{1.45cm}{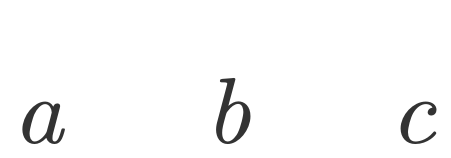}$}
      ~~~\xstep{"Open"}~~~
      \raisebox{-1mm}{$\includepdf{2.35cm}{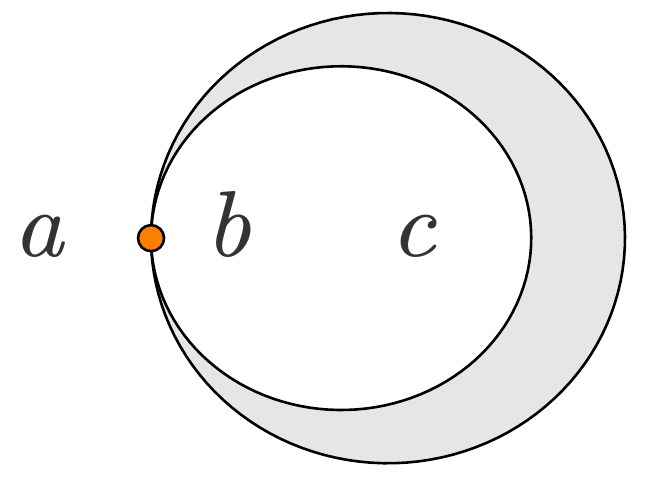}$}
    $
    \caption{Dynamic "proof trace"}
    \label{fig:sharing-assoc-dynamic}
  \end{subfigure}
  \begin{subfigure}[b]{0.48\textwidth}
    \centering
    $\includepdf{3.2cm}{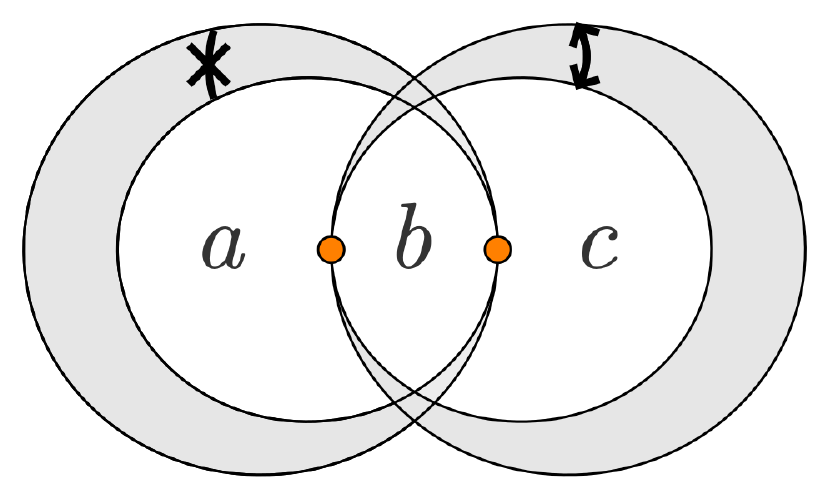}$
    \caption{Static "proof object"}
    \label{fig:sharing-assoc-static}
  \end{subfigure}
  \hfill
  \begin{subfigure}[b]{0.48\textwidth}
    \centering
    $$\stkfig{0.95}{sharing-assoc-graph}$$
    \caption{"Scroll structure with sharing"}
    \label{fig:sharing-assoc-graph}
  \end{subfigure}
  \caption{Example of proof with sharing "inloops"}
  \label{fig:sharing-assoc}
\end{figure}

"Scroll structures" capture Peirce's informal notion of "graph" built out of "atoms" and "scrolls",
and thus the structure of statements in implicative logic. However they cannot express the
\emph{sharing} structure of some proofs in "EGs". Figure~\ref{fig:sharing-assoc-dynamic} shows a
"proof trace" for an analogue of the associativity of conjunction, where grouping of conjuncts is
achieved by enclosing them in "scrolls" with empty "outloops", instead of the traditional "symbolic"
device of parentheses. If one wants to represent faithfully in the corresponding "scroll net" the
identity of the two occurrences of $b$ in the premiss and conclusion, the only way is to have the
"closed" and "opened" "scrolls" overlap on their "inloop", as illustrated in
Figure~\ref{fig:sharing-assoc-static}. Then $b$ has two parents in the "scroll structure"
(Figure~\ref{fig:sharing-assoc-graph}), because the two "inloops" \emph{share} it as a common child.
This means that we need to relax forests into \emph{directed acyclic graphs} (""DAGs""):

\begin{definition}[Sharing]\label{def:sss}
  A ""simple scroll structure with sharing"" $\Scx$ (or \reintro{simple SSS} for short) is the same
  data as a "simple scroll structure", except that the forest becomes a "DAG" with the constraint
  that if a node has more than 1 parent, then all its parents must be "inloops". Formally, if $u
  \Edg v$ and $u' \Edg v$ with $u \not= u'$, then $\exists u_0, u'_0$ such that $u_0 \Att u$ and
  $u'_0 \Att u'$.
\end{definition}

The additional constraint is here justified by the fact that it does not make sense for two distinct
"outloops" to overlap or share content, because the "Open" and "Close" rules only work on "scrolls"
with empty "outloops".

\subparagraph*{Scroll nets}
The last step consists in encoding graph-theoretically the various types of arrows introduced in
Figure~\ref{fig:annotated-illative-transformations} to represent statically "illative
transformations". We again make the distinction between "argumentation" and "interaction" rules,
which act respectively on the nodes and edges of a "SSS":

\newpage

\begin{definition}[Scroll net]\label{def:scroll-net}
  A \reintro{simple scroll net} is a triple $\Snx = \langle \Scx, \Arg, \Int \rangle$ where:
  \begin{itemize}
    \item $\Scx$ is a "simple SSS".
    \item $\intro*\Arg = \langle\intro*\Xjust, \intro*\self\rangle$ is a pair of a directed forest of
    \reintro{justifications} ${\Xjust} \subseteq \Ver_\Scx \times \Ver_\Scx$ and a set of
    \reintro{self-justifications} $\self \subseteq \Ver_\Scx$ called the \reintro{argumentation} of
    $\Snx$. We write respectively $u \just v$ and $\self u$ when $(u, v) \in {\Xjust}$ and $u \in
    {\self}$.
    % , and $\intro*\tjust$ for the reflexive-transitive closure of ${\just}$.
    \item $\intro*\Int = \langle\intro*\Xopen, \intro*\Xclos\rangle$ is a pair of an ""expansion"" and
    a ""collapse"" ${\Xopen}, {\Xclos} \subseteq {\XAtt_\Scx}$ called the \reintro{interaction} of
    $\Snx$. We write $v \intro*\openclos u$ as a shorthand when both $v \open u$ and $v \clos u$.
  \end{itemize}
\end{definition}

We will use letters $\Snx, \Sny, \Snz$ to range over "scroll nets", again subscripting components
accordingly when disambiguation is necessary, so that for instance $v \mathrel{\Xjust_\Snx} u$
expresses that node $v$ justifies node $u$ in $\Snx$. Now, the choice of a directed \emph{forest} of
"justifications" rather than an arbitrary digraph requires some explanations. Compared to an
arbitrary digraph, a forest must satisfy additionally both \emph{acyclicity} and \emph{unicity of
parents}:
\begin{description}
  \item[Unique parents]
  A priori, a node could be the target as well as the source of an arbitrary number of
  "justifications". But upon closer inspection, it appears that while it makes sense to be the
  source of many arrows --- the same statement can "justify" multiple copies of itself, it is
  impossible to be the target of more than one "justification". Indeed in the dynamic understanding
  of "(de)iteration" (Figure~\ref{fig:argumentation-rules}), the "justified" node is either
  "introduced" or "eliminated", and there is no sense in which the exact same node could be
  "introduced" or "eliminated" more than once. Said differently, a node can only be (de)duplicated
  from a single source node.
  \item[Acyclicity]
  Here the formal reasons are much more subtle, although they are also related to the dynamic
  reading of "illative transformations"; we conjecture that the latter indeed preserve acyclicity.
  Intuitively though, it is clear that one wants to avoid cyclic "justifications" in reasoning: there
  is no meaning in asserting that ``$u$ is true because $v$ is true because $u$ is true because $v$
  is true because\ldots''. Acyclicity is also a fundamental component of correctness criterions for
  most "graphical" proof formalisms like proof nets, combinatorial proofs and expansion tree
  proofs~\cite{Miller1987ACR}.
\end{description}

\subparagraph*{Boundaries}

Given a "scroll net", it is not immediately obvious \emph{what} it is proving --- i.e. what are its
\emph{"premiss"} and \emph{"conclusion"}, because both are ``superposed'' in a non-trivial way. For
instance in the "scroll nets" of Figures \ref{fig:modus-ponens-static} and
\ref{fig:sharing-assoc-static}, the same atomic node $b$ occurs both in the "premiss" and the
"conclusion". This is actually a powerful feature, as it makes proofs much more compact by identifying
occurrences that would appear as separate copies in most other proof formalisms, including proof
nets and combinatorial proofs. Fortunately there is a simple algorithm for disentangling "scroll
nets", by exploiting "illative atomicity" and duality. First we recall some standard graph-theoretic
notions:

\begin{definition}
  ~\\
  The \emph{subgraph of $\Snx$ reachable from $v$} is defined as $\intro*\rsubg{v} = \compr{(u,
  w)}{v\tEdg{} u \text{ and } u \Edg w}$. The sets of \emph{parents} and \emph{children} of $v$ are
  defined respectively as $\intro*\prnt{v} = \compr{u}{u \Edg v}$ and $\intro*\chld{v} = \compr{u}{v
  \Edg u}$. We say that $v$ is a ""sibling"" of $u$, written $v \intro*\sibl u$, if $\prnt{v} =
  \prnt{u}$.
\end{definition}

Then we define operations that update a "scroll structure" by adding or removing nodes and edges in
its "DAG":

\begin{definition}
  \begin{itemize}
    \item 
    The ""pruning"" $\intro*\prune{\Scx}{v}$ of a node $v \in \Ver_\Scx$ is the result of
    removing $\rsubg{v}$ and all edges $u_i \Edg v$ from $\Gra_\Scx$, updating other components of
    $\Scx$ accordingly.
    \item
    The ""collapsing"" $\intro*\collapse{\Scx}{v}$ of a "scroll" $v \mathrel{\Att_\Scx} u$ is the
    result of "pruning" every $w \in \chld{v} \setminus \{u\}$, removing $v$, $u$ and their
    associated edges from $\Gra_\Scx$, and adding an edge $v_i \Edg u_j$ for every $v_i \in
    \prnt{v}$ and $u_j \in \chld{u}$.
  \end{itemize}
\end{definition}

We also need to characterize when a node is either \emph{"introduced"}/\emph{"eliminated"} or
\emph{"opened"}/\emph{"closed"} dynamically by some "illative transformation" in the "argumentation"
or "interaction" of a "scroll net". As illustrated in
Figure~\ref{fig:annotated-illative-transformations}, this can be done by just looking at the
\emph{polarity} of nodes:

\begin{definition}[Edit state]
  The sets of \reintro{opened}, \reintro{closed}, \reintro{introduced} and \reintro{eliminated}
  nodes of a "scroll net" $\Snx$ are defined respectively by:
  $$
  \begin{aligned}
    \intro*\OpnNodes{\Snx} &= \compr{v}{\exists u. (v \in \pVer{} \land v \open u) \lor (v \in \nVer{} \land v \clos u)} \\
    \intro*\CloNodes{\Snx} &= \compr{v}{\exists u. (v \in \nVer{} \land v \open u) \lor (v \in \pVer{} \land v \clos u)} \\
    \intro*\InsNodes{\Snx} &= \compr{v}{(v \in \pVer{} \land \exists u. u \just v) \lor (v \in \nVer{} \land \self v)} \\
    \intro*\DelNodes{\Snx} &= \compr{v}{(v \in \nVer{} \land \exists u. u \just v) \lor (v \in
    \pVer{} \land \self v)}
  \end{aligned}
  $$
\end{definition}

We can now define straightforwardly the \emph{"boundaries"} of a "scroll net":

\begin{definition}[Boundaries]
  The ""boundaries"" of a "scroll net" $\Snx$ are two "simple SSS"s:
  \begin{itemize}
    \item the ""premiss"" $\intro*\prem{\Snx} = \collapse{\prune{\Gra_\Snx}{\InsNodes{\Snx}}}{\OpnNodes{\Snx}}$;
    \item the ""conclusion"" $\intro*\conc{\Snx} = \collapse{\prune{\Gra_\Snx}{\DelNodes{\Snx}}}{\CloNodes{\Snx}}$.
  \end{itemize}  
\end{definition}

% \begin{definition}[Boundaries]
%   The \emph{premiss} $\prem{\Snx}$ and \emph{conclusion} $\conc{\Snx}$ of a "scroll net" $\Snx$ are
%   interfaces defined mutually recursively as follows:
%   \small
%   \begin{align*}
%     \conc{\Jx_1, \ldots, \Jx_n} &= \conc{\Jx_1}, \ldots, \conc{\Jx_n} &
%     \prem{\Jx_1, \ldots, \Jx_n} &= \prem{\Jx_1}, \ldots, \prem{\Jx_n} \\
%     \conc{\var{x}{\ax}} = \conc{\jv{y}{x}{\ax}} &= \var{x}{\ax} &
%     \prem{\var{x}{\ax}} = \prem{\bv{x}{\ax}} &= \var{x}{\ax} \\
%     \conc{\bv{x}{\snx}} = \conc{\jbv{y}{x}{\snx}} &= \emptyset &
%     \prem{\jv{y}{x}{\snx}} = \prem{\jbv{y}{x}{\snx}} &= \emptyset \\
%     \conc{\var{x}{\ovscroll{\Snx}{\Sny}}} = \conc{\jv{y}{x}{\ovscroll{\Snx}{\Sny}}} &= \var{x}{\ovscroll{\prem{\Snx}}{\conc{\Sny}}} &
%     \prem{\var{x}{\cvscroll{\Snx}{\Sny}}} = \prem{\bv{x}{\cvscroll{\Snx}{\Sny}}} &= \var{x}{\cvscroll{\conc{\Snx}}{\prem{\Sny}}} \\
%     \conc{\var{x}{\ovcscroll{\Snx}{\Sny}}} = \conc{\jv{y}{x}{\ovcscroll{\Snx}{\Sny}}} &= \prem{\Snx}, \conc{\Sny} &
%     \prem{\var{x}{\cvoscroll{\Snx}{\Sny}}} = \prem{\bv{x}{\cvoscroll{\Snx}{\Sny}}} &= \conc{\Snx}, \prem{\Sny}
%   \end{align*}
% \end{definition}

Note that we apply first "pruning" and then "collapsing" on \emph{sets} of nodes, abstracting from
the particular order in which individual operations are done. We observe that this does not pose any
problem in practice, because this process should always be \emph{confluent}. In fact, we could also
first perform "collapsing" before "pruning" because a node in a "correct" "scroll net" cannot be
both "opened" and "introduced" or both "closed" and "eliminated", i.e. $\OpnNodes{\Snx} \cap
\InsNodes{\Snx} = \emptyset$ and $\CloNodes{\Snx} \cap \DelNodes{\Snx} = \emptyset$. One can easily
convince oneself that the algorithm works by applying it to the various examples of "scroll nets"
presented earlier, where the "premiss" and "conclusion" should match respectively the first and last
"scroll structure" in the corresponding "proof trace".

It is then possible to characterize \emph{"incomplete"} or \emph{partial} "scroll nets" as those
with a non-empty "premiss", just as partial derivations in other formalisms are those where some
leaves are not closed by a nullary rule:

\begin{definition}[Completeness]
  A "scroll net" $\Snx$ is said to be ""complete"" if $\prem{\Snx} = \emptyset$ with $\emptyset$
  denoting the empty "scroll structure" $\langle\langle\emptyset, \emptyset\rangle, \emptyset,
  \emptyset\rangle$, otherwise it is \reintro{incomplete}.
\end{definition}

\section{Correctness}\label{sec:soundness}

\begin{figure}
  \captionsetup[subfigure]{justification=centering}
  \centering
  \begin{subfigure}[b]{0.35\textwidth}
    \centering
    $\includepdf{3.5cm}{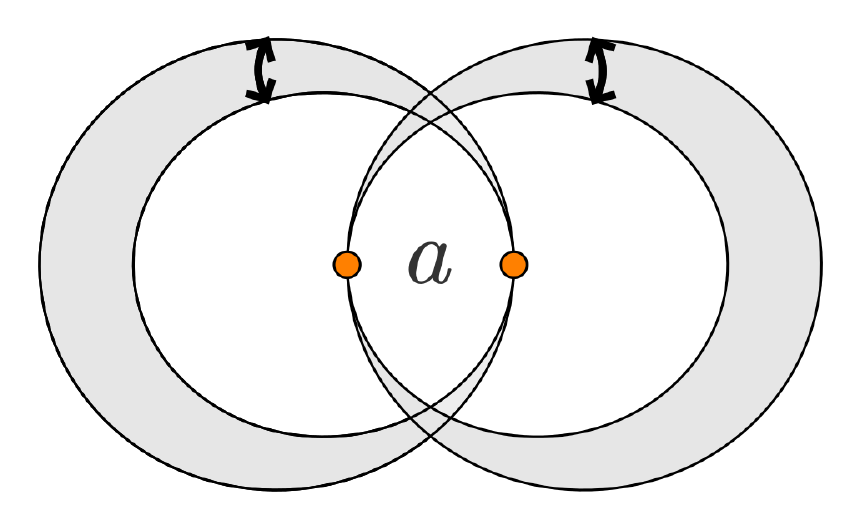}$
    \caption{Overlapping}
    \label{fig:incorrect-overlap}
  \end{subfigure}
  \begin{subfigure}[b]{0.25\textwidth}
    \centering
    $\includepdf{1.5cm}{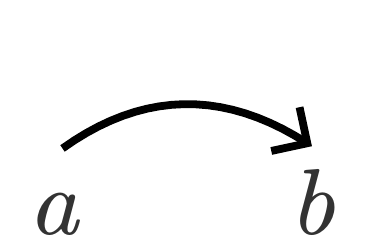}$
    \caption{Identity}
    \label{fig:incorrect-noteq}
  \end{subfigure}
  \hfill
  \begin{subfigure}[b]{0.35\textwidth}
    \centering
    $\includepdf{2.25cm}{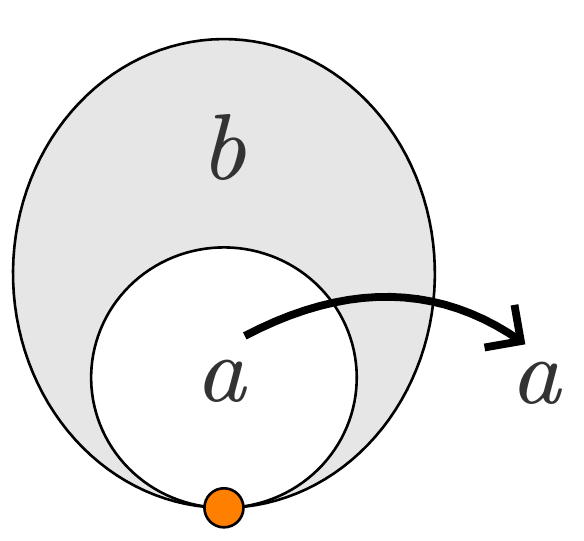}$
    \caption{Scoping}
    \label{fig:incorrect-scope}
  \end{subfigure}
  \caption{Examples of incorrect "scroll nets"}
  \label{fig:incorrect-scroll-nets}
\end{figure}

\subparagraph*{Incorrectness}
In practice, we want the "boundaries" of a "scroll net" to be plain "scroll structures" without
"inloop" sharing, so that they can be interpreted straightforwardly as logical statements. However,
this is only guaranteed for \emph{"correct"} "scroll nets" that can be built by applying a sequence of
"illative transformations", since there is no transformation that makes two "inloops" overlap.

More generally, Definition~\ref{def:scroll-net} is not restrictive enough to capture only the class
of "correct" "scroll nets". Figure~\ref{fig:incorrect-scroll-nets} shows three examples of incorrect
"scroll nets": in Fig.~\ref{fig:incorrect-overlap} it is precisely because of the overlap problem
just mentioned; in Fig.~\ref{fig:incorrect-noteq} it comes from the source and target of a
"justification" not being identical; lastly in Fig.~\ref{fig:incorrect-scope}, it is caused by a
"justification" that violates scoping by going out of a "scroll" that is neither "opened" nor
"closed".

Although there is probably some geometric criterion that could capture each of these mistakes by
analyzing solely the static structure of "scroll nets", in this paper we only formalize the baseline
\emph{sequential} correctness criterion. Informally, we want to say that a "scroll net" $\Snx$ is
"correct" if there exists a total ordering of all elements in its "argumentation" and "interaction"
such that the corresponding dynamic sequence of "illative transformations" builds exactly $\Snx$.

\subparagraph*{Derivations}
To formalize this intuition, we need to define additional operations on "scroll nets" that capture
precisely the \emph{incremental} application of "illative transformations", where the latter are
recorded in the "argumentation" and "interaction". Following standard terminology, we call these
operations ""derivation"" rules and define them in Figure~\ref{fig:derivation-rules}. They are the
fully general version of the more visual examples provided in
Figure~\ref{fig:annotated-illative-transformations}. Although we use the inference line notation,
here we are not building derivation trees: premisses correspond to conditions that the input $\Snx$
must satisfy for the rule to be applicable, while the conclusion $\Snx'$ is the "scroll net" which
results from applying the rule. Thus every rule $\mathsf{R}$ can be seen as a partial map of "scroll
nets". To express the rules as compactly as possible, we rely on the following notational
conventions:
\begin{itemize}
  \item 
  we omit to put $\Snx$ in subscript when referring to its components, and avoid outer brackets $\langle\rangle$ for the triple defining the conclusion $\Snx'$;
  \item
  $X \intro*\dunion Y$ denotes the \emph{disjoint union} of two sets. If $x \in X$ and $y \in Y$,
  then $\intro*\lcpy{x} \in X \dunion Y$ and $\intro*\rcpy{y} \in X \dunion Y$ denote the
  \emph{left} and \emph{right copy} of the original elements; we also write $X \cup x$ as a shorthand for $X \cup \{x\}$;
  \item
  $\Scx \intro*\sssiso \Scx'$ denotes the existence of an \emph{isomorphism} of "scroll structures",
  i.e. an isomorphism between underlying "DAGs"/forests that preserves the "atom" labels and
  "attachments".
  \item
  We write $\intro*\deriv{\Snx}{\Sny}$ to express that there is some rule $\mathsf{R}$ such that
  $\Sny = \mathsf{R}(\Snx)$, and $\intro*\tderiv{\Snx}{\Sny}$ to denote the transitive closure of
  $\Xderiv$, or equivalently that there exists a sequence of rules $\mathsf{R}_1, \ldots,
  \mathsf{R}_n$ such that $\Sny = \mathsf{R}_n \circ \ldots \circ \mathsf{R}_1(\Snx)$.
\end{itemize}

\subparagraph*{Soundness} We now state a series of lemmas whose proofs would ensure the adequacy of
our formal definition of "derivation" rules. Again for lack of space, we omit such proofs. First we
want to ensure that if $\Snx$ is a "scroll net" and $\deriv{\Snx}{\Sny}$, then $\Sny$ is still a
well-formed "scroll net":

\begin{lemma}[Well-formedness preservation]
  If $\deriv{\Snx}{\Sny}$ then the following holds:
  \begin{itemize}
    \item $\Gra_\Sny$ is a "DAG" satisfying the constraint of Definition~\ref{def:sss};
    \item ${\XAtt_\Sny} \subseteq {\XEdg_\Sny}$ and it satisfies the well-formedness conditions of Definition~\ref{def:scroll-structure};
    \item $\Xjust_\Sny$ is a forest.
  \end{itemize}
\end{lemma}

Then we want to ensure that "derivation" rules do not change the "boundaries" of a "scroll net", which is akin to \emph{subject reduction} in type theory:

\begin{definition}
  We say that $\Snx$ is ""interpretable"" iff $\Gra_{\prem{\Snx}}$ and $\Gra_{\conc{\Snx}}$ are forests.
\end{definition}

\begin{lemma}[Subject construction]
  If $\deriv{\Snx}{\Sny}$ then the following holds:
  \begin{itemize}
    \item\textbf{(Premiss preservation)} $\prem{\Snx} \sssiso \prem{\Sny}$.
    \item\textbf{(Interpretability)} if $\Snx$ is "interpretable", so is $\Sny$.
  \end{itemize}
\end{lemma}

We can then show the following results about logical soundness:

\begin{definition}[Interpretation]
  The interpretation $\intro*\interp{\Scx}$ of a "scroll structure" $\Scx$ is a formula defined by
  induction on the depth of $\Gra_\Scx$:
  $$
  \small
  \begin{aligned}
    &\interp{\emptyset} = \top \qquad \interp{v_1, \ldots, v_n} = \interp{v_1} \land \ldots \land \interp{v_n} \\
    &\interp{v} = \begin{cases}
      a &\text{if $\chld{v} = \emptyset$ and $\Lab(v) = a$} \\
      \interp{v_1, \ldots, v_n} \limp \interp{w_1, \ldots, w_m} &\text{if $\chld{v} = \{v_1, \ldots, v_n, u\}$, $u \Att w$ and $\chld{w} = \{w_1, \ldots, w_m\}$}
    \end{cases} \\
  \end{aligned}
  $$
  We write $\intro*\pinterp{\Snx}$ and $\intro*\cinterp{\Snx}$ as a shorthand for
  $\interp{\prem{\Snx}}$ and $\interp{\conc{\Snx}}$.
\end{definition}

\begin{lemma}
  If $\Scx \sssiso \Scy$ then $\interp{\Scx} \lequiv \interp{\Scy}$, i.e. $\interp{\Scx}$ and $\interp{\Scy}$ are logically equivalent.
\end{lemma}

\begin{lemma}[Conclusion entailment]
  If $\deriv{\Snx}{\Sny}$ and $\Snx$ is "interpretable", then $\cinterp{\Snx} \vdash \cinterp{\Sny}$.
\end{lemma}

\begin{theorem}[Soundness]
  If $\tderiv{\Scx}{\Snx}$ then $\interp{\Scx} \vdash \cinterp{\Snx}$.
\end{theorem}

\begin{corollary}
  If $\tderiv{\emptyset}{\Snx}$ then $\vdash \cinterp{\Snx}$.
\end{corollary}

Finally, the above soundness results legitimate our sequential definition of correctness:

\begin{definition}[Correctness]
  We say that a "scroll net" $\Snx$ is ""correct"" iff $\Snx$ is "interpretable" and
  $\tderiv{\prem{\Snx}}{\Snx}$.
\end{definition}

\subparagraph*{Composition}
A first very natural way to compose two "scroll nets" $\Snx$ and $\Sny$ is by taking their
\emph{"juxtaposition"}. As usual with "graphical" formalisms, this is defined by taking the
\emph{disjoint union} of their respective components:

\begin{definition}[Horizontal composition]
  The ""horizontal composition"" of two "scroll structures" $\Scx$ and $\Scy$ and two "scroll nets" $\Snx$ and $\Sny$ are defined by
  $$
  \begin{aligned}
  \Scx \dunion \Scy &= \langle \Gra_\Scx \dunion \Gra_\Scy, \Lab_\Scx \dunion \Lab_\Scy, \XAtt_\Scx \dunion \XAtt_\Scy \rangle \\
  \Snx \dunion \Sny &= \langle\Scx_\Snx \dunion \Scx_\Sny, \langle{\Xjust_\Snx} \dunion {\Xjust_\Sny}, \self_\Snx \dunion \self_\Sny\rangle, \langle{\Xopen_\Snx} \dunion {\Xopen_\Sny}, {\Xclos_\Snx} \dunion {\Xclos_\Sny}\rangle\rangle
  \end{aligned}
  $$
\end{definition}

The "boundaries" of $\Snx \dunion \Sny$ will be interpreted as the \emph{conjunction} of the
"boundaries" of $\Snx$ and $\Sny$, i.e. $\pinterp{\Snx \dunion \Sny} \lequiv \pinterp{\Snx} \land
\pinterp{\Sny}$ and $\cinterp{\Snx \dunion \Sny} \lequiv \cinterp{\Snx} \land \cinterp{\Sny}$.
Importantly, it should also be the case that $\Snx \dunion \Sny$ is "correct" whenever $\Snx$ and
$\Sny$ are. Indeed, the intuition is that the two "derivations" $\tderiv{\prem{\Snx}}{\conc{\Snx}}$
and $\tderiv{\prem{\Sny}}{\conc{\Sny}}$ can be performed in parallel to yield a "derivation"
$\tderiv{\prem{\Snx} \dunion \prem{\Sny}}{\conc{\Snx} \dunion \conc{\Sny}}$. This kind of
composition is typical of \emph{deep inference} formalisms like the calculus of structures or open
deduction \cite{tubella:hal-02390267}.

The more standard kind of composition corresponds to the cut rule in sequent calculus or to the composition of morphisms in categorical semantics, and we call it \emph{"vertical composition"}.
% For now we restrict its domain to "correct" "scroll nets":

\begin{definition}[Compatibility]
  We say that two "scroll nets" $\Snx$ and $\Sny$ are ""compatible"", written $\Snx \intro*\compat
  \Sny$, whenever $\conc{\Snx} \sssiso \prem{\Sny}$.
\end{definition}

\begin{definition}[Superposition]\label{def:superposition}
  The ""superposition"" $\Snx \intro*\superp \Sny$ of two "correct" and "compatible" "scroll nets"
  $\Snx \compat \Sny$ is defined as the lifting of the "derivation" $\tderiv{\conc{\Snx} \sssiso
  \prem{\Sny}}{\Sny}$ into a "derivation" $\tderiv{\Snx}{\Snx \superp \Sny}$, i.e. the same sequence
  of rules is applied modulo the isomorphism $\conc{\Snx} \sssiso \prem{\Sny}$.
\end{definition}

\begin{definition}[Vertical composition]
  Given two "correct" and "compatible" "scroll nets" $\Snx \compat \Sny$, their ""vertical
  composition"" $\Sny \circ \Snx$ is defined by taking the respective "derivations"
  $\tderiv{\prem{\Snx}}{\Snx}$ and $\tderiv{\prem{\Sny}}{\Sny}$, and composing those into a
  "derivation" $\tderiv{\prem{\Snx}}{\Snx \superp \Sny = \Sny \circ \Snx}$.
\end{definition}

These definitions rely on the fact that one can choose \emph{canonically} a "derivation" for a
"correct" "scroll net" $\Snx$, corresponding to our aforementioned intuition that there should
always be a way to totally order the elements of $\Arg_\Snx \dunion \Int_\Snx$ into the associated
sequence of "illative transformations". It is also not entirely clear that the so-called
\emph{"superposition"} operation is formally well defined. The dynamic intuition is that the nodes
and edges that appear in $\conc{\Snx}$ are included in $\Snx$, so that every rule applicable in
$\conc{\Snx}$ can equally be applied in $\Snx$ (what we called ``lifting'' in
Definition~\ref{def:superposition}). The static, geometric intuition is that $\Snx$ and $\Sny$ being
"compatible" means that they have a common (modulo isomorphism) boundary, so that one can literally
superpose the topological representations of $\Snx$ and $\Sny$ on this boundary to obtain $\Snx
\superp \Sny$. This can be visualized in Figures~\ref{fig:modus-ponens} and \ref{fig:sharing-assoc}
by turning the "proof traces" into recording "derivations", splitting said "derivations" into
sub-"derivations", and checking that their recombinations through "superposition" gives back the
expected "scroll net".

\section{Computation}\label{sec:computation}

\subparagraph*{Detours}
In natural deduction, a so-called \emph{detour} --- corresponding to a \emph{$\beta$-redex} in
simply typed $\lambda$-calculus --- arises when an introduction rule on some formula is followed by
an elimination rule on the same formula. Similarly in "scroll nets", we will call ""detour"" a node
that is both "introduced" and "eliminated", in the sense of being entirely scribed and then entirely
erased from $\SA$. A simple argument by combinatorial exhaustion shows that there are only 4
possible shapes of "detours", depicted on the left-hand side in Figure~\ref{fig:detours}. Indeed, a
"detour" is always a "scroll" that falls under one of the following cases:
\begin{description}
  \item[Interaction/Interaction ($\intro*\red{ii}$)] "opened" then "closed", either in a "positive" or
  "negative" location;
  \item[Interaction/Argumentation ($\red{ia}$)] "opened" then "deleted", or "inserted" then
  "closed";
  \item[Argumentation/Interaction ($\red{ai}$)] "iterated" then "closed", or "opened" then
  "deiterated";
  \item[Argumentation/Argumentation ($\red{aa}$)] "iterated" then "deleted", or "inserted" then
  "deiterated".
\end{description}
Although we have given 8 cases, they can be divided in 4 pairs that have the same shape: it is then
the polarity of the "detour" that determines in which order the two "illative transformations" have
been performed. There is also a 9\textsuperscript{th} case when two "argumentation" rules interact
on an "atom", but we consider it as a variant of the previous "argumentation"/"argumentation" case
to preserve symmetry.

\subparagraph*{Detour reduction}
In Figure~\ref{fig:detours}, we give for each kind of "detour" a general \emph{reduction rule}
(subdivided in a "scroll" and "atom" case for $\red{aa}$). Here we depict the "detour" in a
"positive" area, but this also works by inverting polarities thanks to the aforementioned
polarity-invariance. Currently these rules are experimental, although one can check that they
correctly preserve the "boundaries" of the "scroll net". To do so, it is necessary to observe the
two following facts:
\begin{itemize}
  \item the "premiss" and "conclusion" become inverted when inverting the polarity. This means that
  it is the "premiss" (resp. "conclusion") of the antecedant of a "scroll" --- which is itself a
  "scroll net" --- that appears in the "conclusion" (resp. "premiss") of the "scroll";
  \item for any two composable "scroll nets" $\Snx$ and $\Sny$, $\prem{\Snx \superp \Sny} =
  \prem{\Snx}$ and $\conc{\Snx \superp \Sny} = \conc{\Sny}$.
\end{itemize}
Note that we depict a "scroll net" $\Snx$ that appears inside another "scroll net" $\Sny$ --- i.e. a
\emph{subnet} of $\Sny$ --- by enclosing it in a two-part box labelled $\Snx$ on the right, where
the upper and lower parts contain respectively $\prem{\Snx}$ and $\conc{\Snx}$.

\subparagraph*{Simulating STLC}
Given the above "detour" reduction rules, it is now possible to simulate straightforwardly the
simply typed $\lambda$-calculus. We express this translation diagrammatically in
Figure~\ref{fig:simul}, awaiting future work for a more rigorous graph-theoretic formalization. The
translation is divided into two parts:
\begin{description}
  \item[Static typing (Figure~\ref{fig:simul-typing-rules})]
  Each typing rule with premisses $\Gamma_i \vdash t_i : A_i$ for $1 \leq i \leq n$ and conclusion
  $\Gamma \vdash t : A$ is mapped to a "scroll net" $\Snx$ such that $\prem{\Snx}$ (resp.
  $\conc{\Snx}$) is equal to (the translation of) $\Gamma$ (resp. $A$), and where the inductive
  translations of sub-derivations appear as subnets $\Snx_i$ with corresponding "boundaries"
  $\prem{\Snx_i} = \Gamma_i$ and $\conc{\Snx_i} = A_i$.
  \item[Dynamic computation (Figure~\ref{fig:simul-beta-reduction})]
  Here we simulate the $\beta$-reduction rule with the two "detour" reduction rules $\red{aa}$ and
  $\red{ii}$. Note that this does not trigger any form of duplication or erasure, as would be the
  case with substitution in $\lambda$-calculus. We believe that substitution should happen when new
  "detours" in $u$ and $t$ generated by the application of the previous two rules are further
  reduced.
\end{description}

% \begin{table}[]
%   \centering
%   \def\arraystretch{1.4}%
%   \begin{tabular}{|l|c|c|}
%     \hline
%      & \textsf{i}\textbf{nteraction} & \textsf{a}\textbf{rgumentation} \\
%     \hline
%     \textsf{i}\textbf{nteraction}    &  $\var{x}{\ocscroll{}{\Snx}}$  &  $\bv{x}{\oscroll{\ascnet{\Snx}{\Ix}{}}{\ascnet{\Sny}{}{\Iy}}}$  \\
%     \hline
%     \textsf{a}\textbf{rgumentation} &  $\jv{y}{x}{\cscroll{\ascnet{\Snx}{}{\Ix}}{\ascnet{\Sny}{\Iy}{}}}$  &  $\jbv{y}{x}{\scroll{\Snx}{\Sny}} \ \ \text{or}\ \ \jbv{y}{x}{a}$ \\
%     \hline
%   \end{tabular}
%   \caption{The 4 types of detours}
%   \label{fig:detours}
% \end{table}

% \section{Simulating STLC}\label{sec:stlc}

\section{Conclusion}\label{sec:conclusion}

In this article we have laid out the foundations of the theory of "scroll nets", focusing on its
historical and conceptual genesis, its technical graph-theoretic formalization, and sketching its
connection to the simply typed $\lambda$-calculus. As mentioned repeatedly, there is much room for
improvement and development of the meta-theory of our formalism. In particular, we wish to find
mathematical proofs for the various \emph{correctness} results, as well as a more rigorous
definition of the \emph{"superposition"} operation $\superp$. The latter should benefit from a
\emph{sequentialization theorem} formalizing the existence of a total ordering on the "illative
transformations" of a "scroll net". A deeper analysis of "detours" and "detour" elimination would also
require an entire dedicated article. In the remainder, we summarize connections that the theory of
"scroll nets" entertains with previous, related, and future works and applications that we envision.

% \begin{remark}
%   When judgments become identified with statements, \emph{""proof states""} become the same as
%   \emph{proof statements}! This very much concords with the etymology of the word ``statement''
%   which is the act of \emph{stating}, that is saying something about the \emph{state} of the world.
% \end{remark}

\subparagraph*{Intuitionistic "EGs"}
In previous work \cite{flower-calculus}, we have shown how to capture precisely provability in
(full) intuitionistic first-order logic inside a variant of Oostra's system of intuitionistic "EGs"
\cite{oostra_graficos_2010} dubbed \emph{flower calculus}. Although it enjoys a nice metaphorical
notation and some useful properties in the context of "interactive theorem proving" --- such as
\emph{analyticity} and \emph{invertibility} of all inference rules, it breaks the perfect duality
stemming from the "illative atomicity" of rules found in the original approach of Peirce and Oostra,
which was shown in this paper to be essential to the application of the Curry-Howard methodology to
"EGs".

\subparagraph*{Generalized "scroll"}
In \cite{oostra_graficos_2010}, Oostra introduces a \emph{horizontal} generalization of the "scroll"
where it can have an arbitrary number $n$ of "inloops", subsuming "seps" as the case where $n = 0$.
In our formalization, this would correspond to the possibility of having $n$ \emph{"sibling"}
"inloops" $v_1 \sibl \ldots \sibl v_n$ in the same "scroll" $u$, i.e. $u \Att v_1 \ldots v_n$.
Following the classical reading of the "scroll" as nested negations, one naturally interprets this
construct through De Morgan equivalences by taking the \emph{disjunction} of "inloops", and the
aforementioned works show how one can still capture intuitionistic logic in this setting. We have
observed in preliminary work that one can also consider a \emph{vertical} generalization by allowing
chains of "attachments" of the form $v_1 \Att \ldots \Att v_m$, leading to a notion of
$(n,m)$-"scroll". Chains of "attachments" are naturally interpreted as \emph{intuitionistic
subtraction}, which should enable a novel treatment of \emph{dual-intuitionistic} and
\emph{bi-intuitionistic} logic.

\subparagraph*{Classical logic}
Peirce's original system of "EGs" for propositional logic was called \sys{Alpha}, and it captured
\emph{classical} logic by adopting the classical reading of the "scroll". A straightforward way to
adapt "scroll nets" to this setting would consist in dropping the "attachments" $\XAtt$ in
Definition~\ref{def:scroll-structure}, just as Peirce ignored them. However since "seps" can be seen
as $(0,0)$-"scrolls", we believe that a more general treatment would keep the notion of "attachment",
and instead characterize classical proofs as those where "(de)iterations" do not preserve
\emph{continuity}; that is, where an attached "inloop" can be duplicated into an unattached "sep".
Once the right framing of classical logic has been found, we believe it could provide interesting
insights into the problem of finding a good notion of \emph{proof identity} in classical logic
\cite{strasburger-problem-2019}, as well as a decomposition of the computational behavior of
classical systems in the Curry-Howard tradition like the $\lambda\mu$-calculus
\cite{parigotLmCalculusAlgorithmicInterpretation1992} and System \sys{L}
\cite{munch-maccagnoniFocalisationClassicalRealisability2009}.

\subparagraph*{Other logics}
Peirce had also devised extensions of \sys{Alpha} that capture \emph{first-order predicate} logic
(\sys{Beta}) and somewhat more speculatively \emph{modal} and \emph{higher-order} logics
(\sys{Gamma}), thus providing natural venues for extensions of "scroll nets" to these more
expressive logics. Our discovery of the computational content of "EGs" and its close connection to
$\lambda$-calculus should enable a \emph{type-theoretic} approach to modal and higher-order logics,
hopefully shedding light on both theory and applications to programming and "interactive theorem
proving".

\subparagraph*{Combinatorial proofs}
Intuitionistic combinatorial proofs have been recently introduced by Heijltjes et al. as ``a
concrete geometric semantics of intuitionistic logic'' \cite{heijltjes_intuitionistic_2019}. They
also benefit from a computational interpretation \cite{heijltjesNormalizationSyntax2022} and have
been related to game semantics \cite{heijltjes_intuitionistic_2019}. However they exhibit a quite
different graph-theoretic structure, with a strict separation between formulas in the arena and
formulas in the game. In contrast, nodes in the "scroll structure" and "argumentation"/"interaction"
of a "scroll net" are shared, leading to a more compact representation. Like "scroll structures",
arenas identify formulas that are equivalent modulo commutativity and associativity of conjunction.
But contrary to "scroll structures", they also conflate formulas equivalent modulo currying, which
makes them unable to express computations under nested abstractions.

% \subparagraph*{Proof complexity}
% Argue that "scroll nets" are a good candidate for a proof complexity framework, because they give a very compact representation of proofs that maximizes sharing of statements occurrences and of subproofs.

\subparagraph*{Bigraphs}
"Scroll nets" are closely related to the notion of \emph{bigraph} introduced by Milner as a model of
mobile interaction \cite{milnerBigraphicalReactiveSystems2001}. A bigraph consists of two
independent structures: a \emph{topograph} or \emph{place graph} encoding spatial information as a
forest, and a \emph{monograph} or \emph{link graph} encoding the connectivity of nodes as a
hypergraph sharing the same vertices as the topograph. They have been generalized along two
independent axes: the \emph{bigraphs with sharing} of \cite{sevegnaniBigraphsSharing2015} relax the
topograph forests into "DAGs" to represent overlapping of locations; and the \emph{directed bigraphs}
of \cite{grohmannDirectedBigraphs2007} orient the monograph's edges to encode resource dependencies
or information flow. "Scroll structures" have the same structure as topographs in bigraphs with
sharing, while (static) "illative transformations" seem to be special cases of monographs in
directed bigraphs. It is remarkable that "scroll nets" combine these two generalizations of a very
recent model of computation, while being based entirely on the principles of "EGs" that predate both
proof theory and the advent of computer science. It would be interesting to explore the possibility
of formalizing "illative transformations" and "detour" reduction rules as a \emph{bigraphical reactive
system}, possibly in a categorical setting such as adhesive categories
\cite{lackAdhesiveQuasiadhesiveCategories2005}.

% \subparagraph*{Cyclic proofs}

% \subparagraph*{Denotational semantics}
% Coming back (again) to the question of identity of proofs: could the detour reduction rules provide
% a novel equational theory on proofs, with associated category-theoretic axiomatizations?

\subparagraph{Proof script vs proof term}
We mentioned in the introduction the distinction between the notions of "proof script" and "proof term"
in the interface of state-of-the-art "ITPs" like Rocq and Lean, which we coined more generally as
that between "proof objects" and "proof traces". This distinction seems to be essential to their
successful usage in large-scale formalization efforts, but is unfortunately not reflected in current
proof-theoretical frameworks. While "proof term" languages are based on dependent type theory and thus
benefit from strong and (relatively) uniform theoretical foundations, "proof scripts" are expressed in
a variety of metaprogramming and domain-specific languages with somewhat ad hoc structure and
semantics\footnote{We only know of one attempt to give rigorous denotational semantics to tactic
languages for proof refinement \cite{sterling_algebraic_2017}.}. This has led to a fragmented
landscape of user interfaces for "ITPs" that limits their accessibility and interoperability, as
well as the common belief that there is unavoidable accidental complexity in the various processes
(e.g. type inference, proof search, elaboration, macro processing, pretty-printing) that bridge the
gap between low-level terms and high-level tactics/notations.

We believe that "scroll nets" could provide a unified foundation for representing both "proof scripts"
and "proof terms", without conflating the two notions. The reason is that "illative transformations"
can be seen both as proof refinement primitives when used dynamically as "derivation" rules, and as a
compact/parallel representation of the information flow in "proof objects" when recorded statically
inside "scroll nets". This could allow for a more principled, systematic study of the properties of
tactics and their interaction with "proof terms", in a way that is not possible with current
approaches.

\bibliography{main}

\appendix

\begin{figure}[h]
  \begin{framed}
  \footnotesize
  \begin{mathpar}
    \inferrule[Open$+$]
      {v_1, \ldots, v_n \in \pVer{} \and \forall i,j \text{ such that } 1 \leq i,j \leq n \text{ and } i \not= j,\  v_i \mathrel{\Xsibl_{\conc{\Snx}}} v_j \and u, u' \not\in \Ver}
      {
        \left\langle\left\langle \Ver \cup \{u,u'\},
        \text{$\begin{array}{rl}
          {\XEdg}\!\!\!\!\!\! &\setminus \compr{(w,v_i)}{1 \leq i \leq n \land w \Edg v_i} \\
                             &\cup      \compr{(w,u)}{\exists i. w \Edg v_i} \\
                             &\cup      \,(u,u') \\
                             &\cup      \compr{(u',v_i)}{1 \leq i \leq n}
        \end{array}$}
        \right\rangle, \Lab, {\XAtt} \cup (u,u')
        \right\rangle, \Arg, \langle {\Xopen} \cup (u,u'), {\Xclos} \rangle
      }
    \and
    \inferrule[Open$-$]
      {v_1, \ldots, v_n \in \nVer{} \and \forall i,j \text{ such that } 1 \leq i,j \leq n \text{ and } i \not= j,\  v_i \mathrel{\Xsibl_{\conc{\Snx}}} v_j \and u, u' \not\in \Ver}
      {
        \left\langle\left\langle \Ver \cup \{u,u'\},
        \text{$\begin{array}{rl}
          {\XEdg}\!\!\!\!\!\! &\setminus \compr{(w,v_i)}{1 \leq i \leq n \land w \Edg v_i} \\
                             &\cup      \compr{(w,u)}{\exists i. w \Edg v_i} \\
                             &\cup      \,(u,u') \\
                             &\cup      \compr{(u',v_i)}{1 \leq i \leq n}
        \end{array}$}
        \right\rangle, \Lab, {\XAtt} \cup (u,u')
        \right\rangle, \Arg, \langle {\Xopen}, {\Xclos} \cup (u,u') \rangle
      }
    \and
    \inferrule[Close$+$]
      {v \in \pVer{} \and v \Att u \and (v,u) \not\in {\Xclos}}
      {\Scx, \Arg, \left( \Xopen, {\Xclos} \cup (v,u) \right)}
    \and
    \inferrule[Close$-$]
      {v \in \nVer{} \and v \Att u \and (v,u) \not\in {\Xopen}}
      {\Scx, \Arg, \left( {\Xopen} \cup (v,u), {\Xclos} \right)}
    \and
    \inferrule[Insert]
      {v \in \pVer{} \setminus \dom(\Lab) \and \text{$\phi$ is a "scroll structure" with a single root $u$}}
      {\langle\langle \Ver \dunion \Ver_\phi, (\XEdg \dunion \XEdg_\phi) \cup (\lcpy{v}, \rcpy{u})\rangle, \Lab \dunion \Lab_\phi, \XAtt \dunion \XAtt_\phi\rangle, (\Xjust, \self \cup \rcpy{u}), \Int}
    \and
    \inferrule[Delete]
      {v \in \pVer{} \and v \not\in \self}
      {\Scx, \langle\Xjust, \self \cup v\rangle, \Int}
    \and
    \inferrule[IterateRoot]
      {\prnt{u} = \emptyset}
      {\left\langle\left\langle \Ver \dunion \Ver_{\conc{\rsubg{u}}}, {\XEdg} \dunion {\XEdg_{\conc{\rsubg{u}}}} \right\rangle, \Lab \dunion \Lab_{\conc{\rsubg{u}}}, \XAtt \dunion \XAtt_{\conc{\rsubg{u}}} \right\rangle, \langle {\Xjust} \cup (\lcpy{u}, \rcpy{u}), \self \rangle, \Int}
    \and
    \inferrule[IterateDeep]
      {v \in \nVer{} \setminus \dom(\Lab) \and \neg\exists u_0.\, u_0 \Att u \and \exists v_0.\, u \sibl v_0 \land v_0 \tEdg{} v}
      {\left\langle\left\langle \Ver \dunion \Ver_{\conc{\rsubg{u}}}, {\XEdg} \dunion {\XEdg_{\conc{\rsubg{u}}}} \cup (\lcpy{v},\rcpy{u}) \right\rangle, \Lab \dunion \Lab_{\conc{\rsubg{u}}}, \XAtt \dunion \XAtt_{\conc{\rsubg{u}}} \right\rangle, \langle {\Xjust} \cup (\lcpy{u}, \rcpy{u}), \self \rangle, \Int}
    \and
    \inferrule[Deiterate]
      {v \in \nVer{} \and \neg\exists u_0.\, u_0 \Att u \and \exists v_0.\, u \sibl v_0 \land v_0 \tEdg{} v \and \conc{\rsubg{u}} \sssiso \prem{\rsubg{v}}}
      {\Scx, \langle {\Xjust} \cup (u,v), \self \rangle, \Int}
  \end{mathpar}
  \end{framed}
  \caption{"Derivation" rules for "scroll nets"}
  \label{fig:derivation-rules}
\end{figure}

\begin{figure}
  $$
  \begin{array}{r@{\quad}c@{\quad}l}
    \includepdf{4cm}{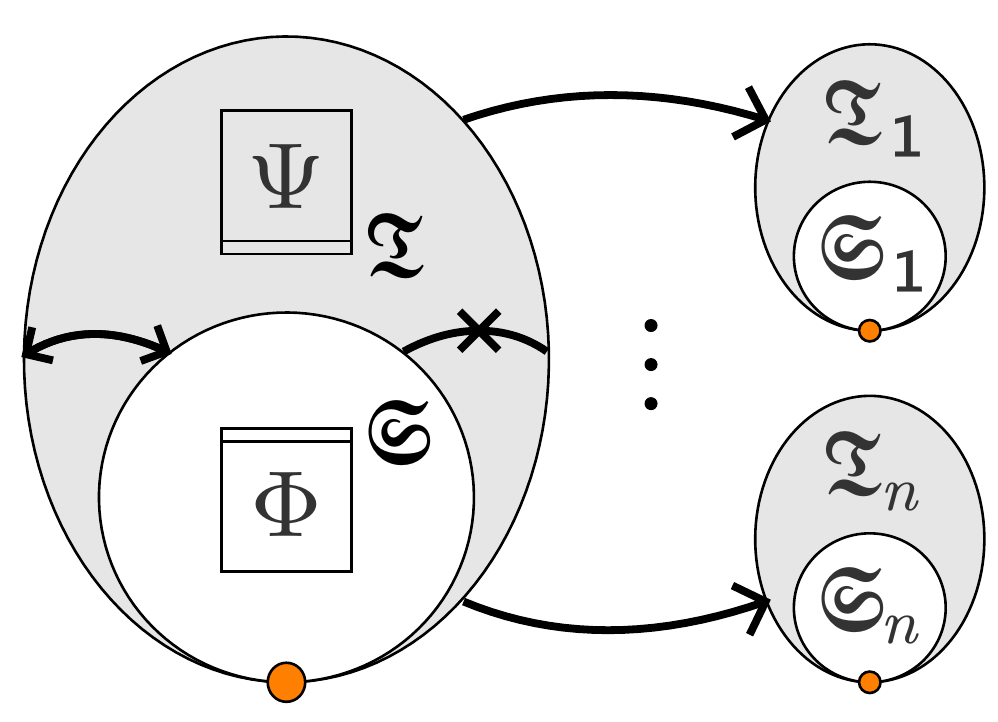} &\red{ii}& \includepdf{1.7cm}{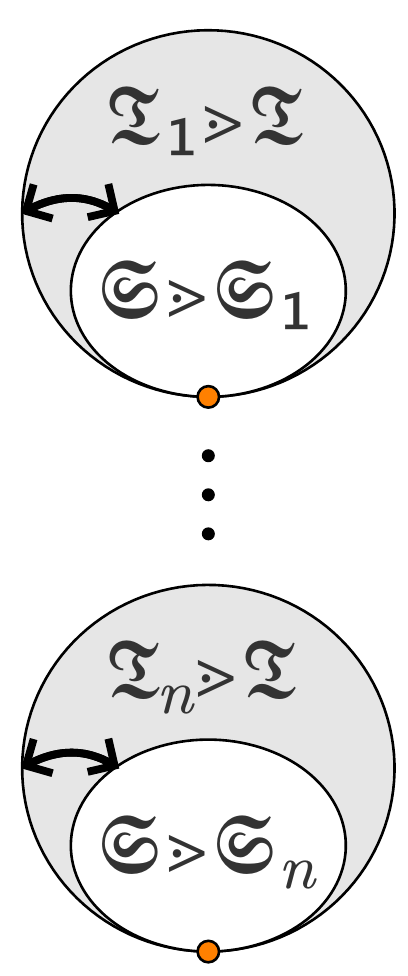} \\[2cm]
    \includepdf{4cm}{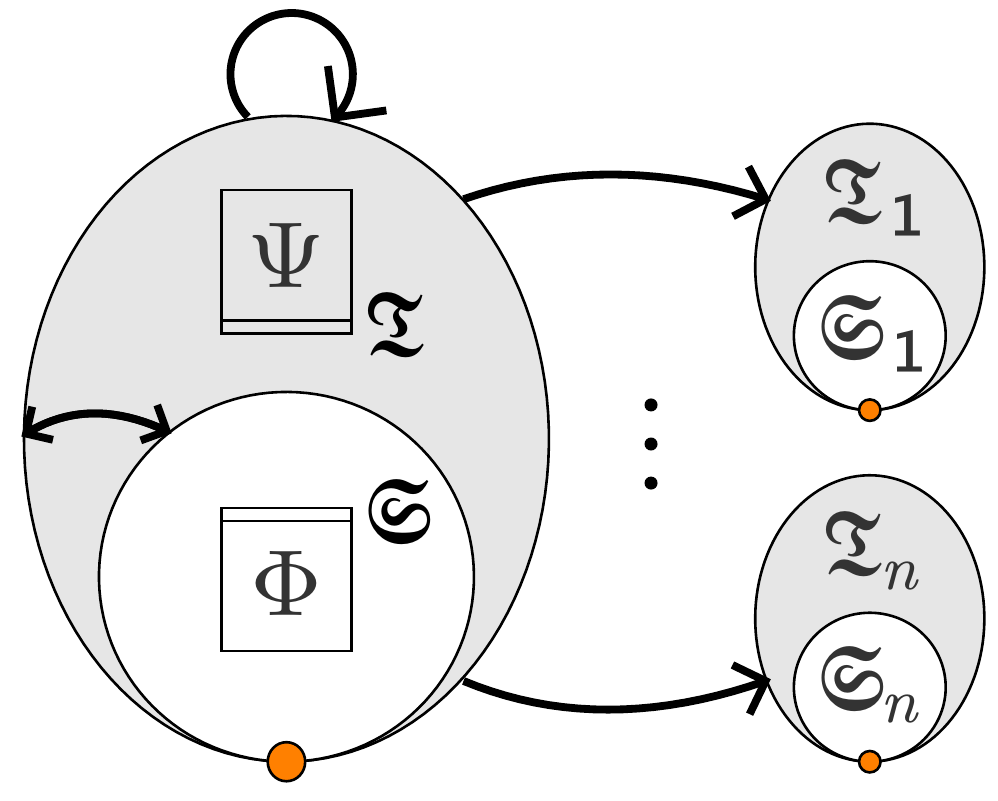} &\red{ia}& \includepdf{1.7cm}{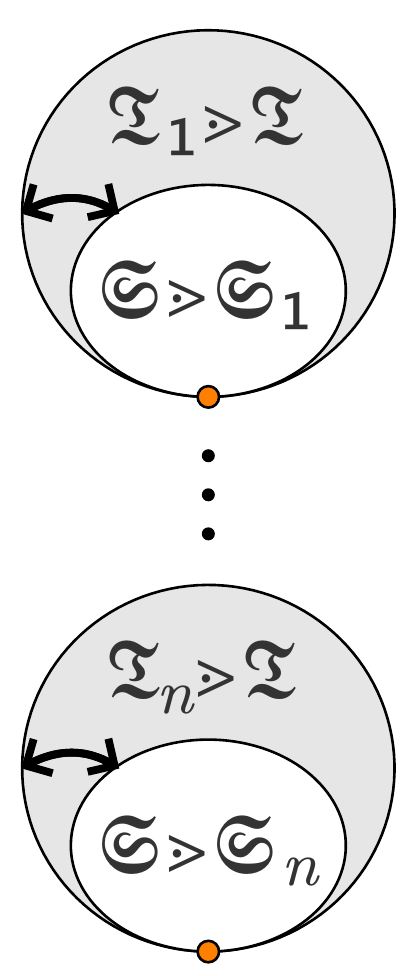} \\[2cm]
    \includepdf{4.65cm}{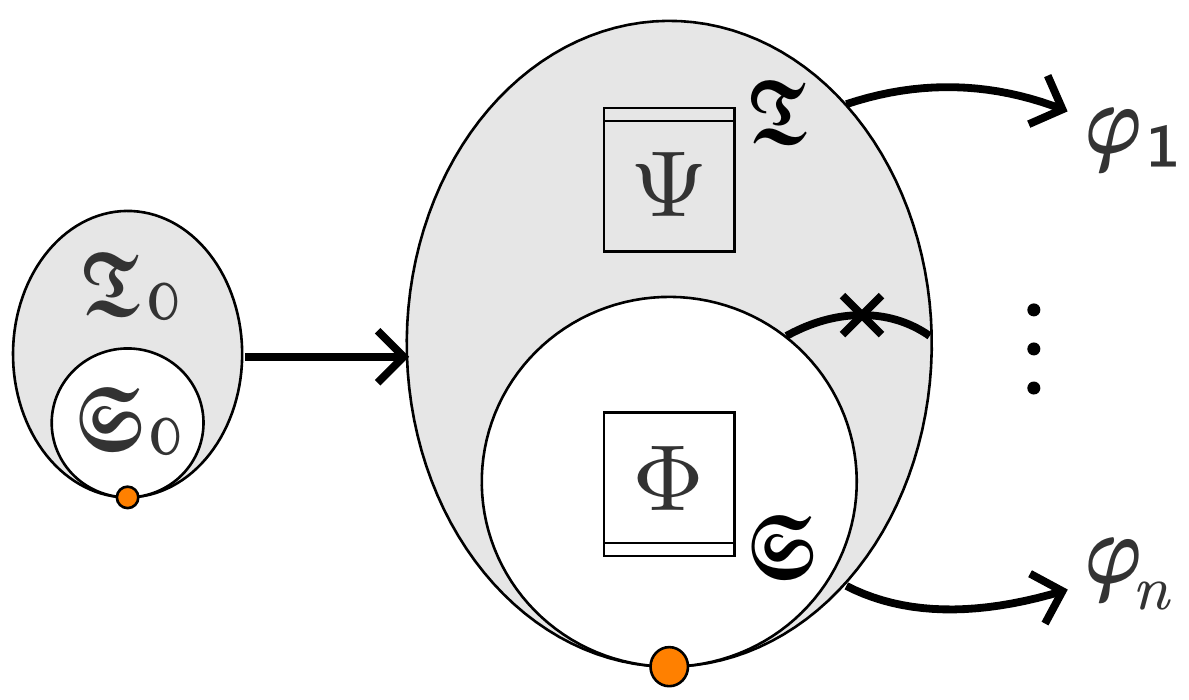} &\red{ai}& \includepdf{4.65cm}{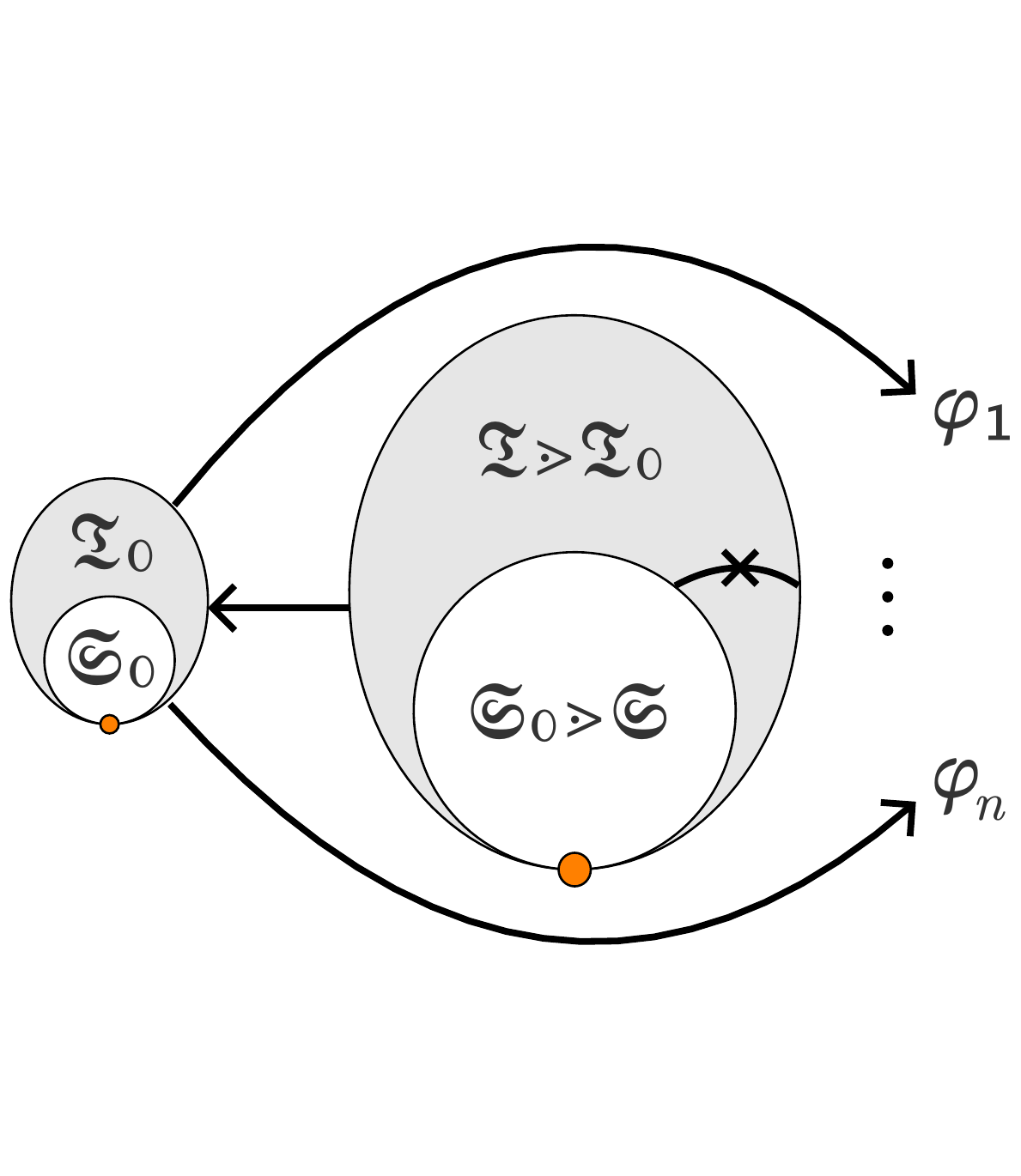} \\[2cm]
    \includepdf{4.65cm}{aas1} &\red{aa}& \includepdf{2.8cm}{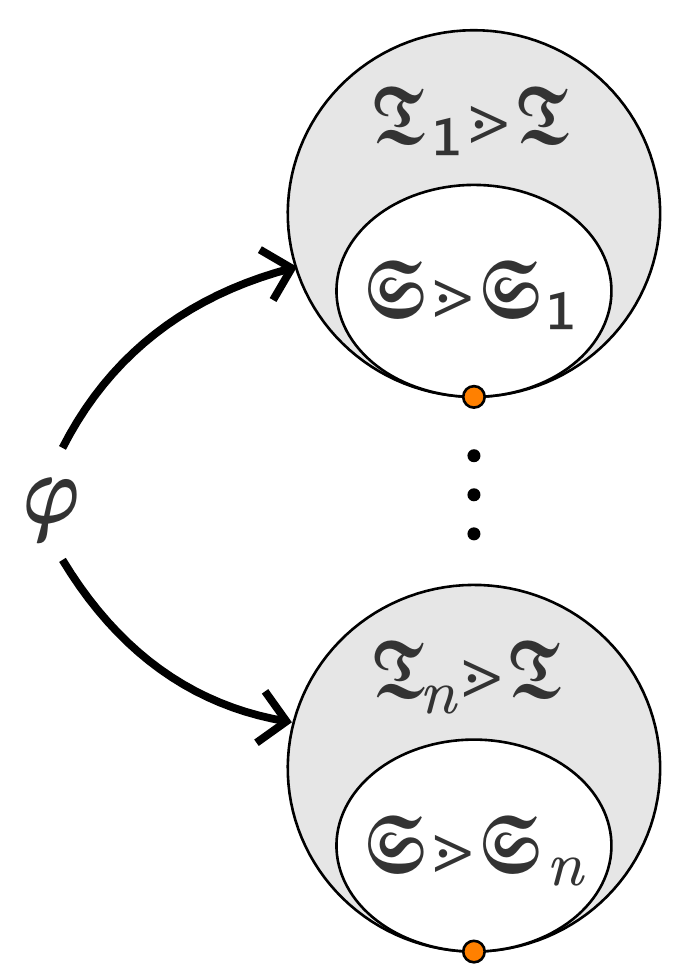} \\[2cm]
    \includepdf{2.5cm}{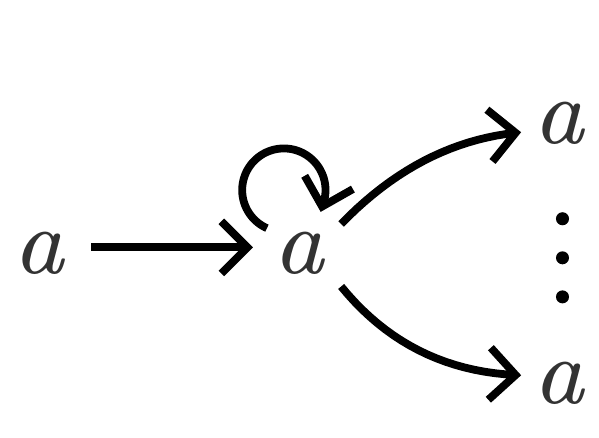} &\red{aa}& \includepdf{2.5cm}{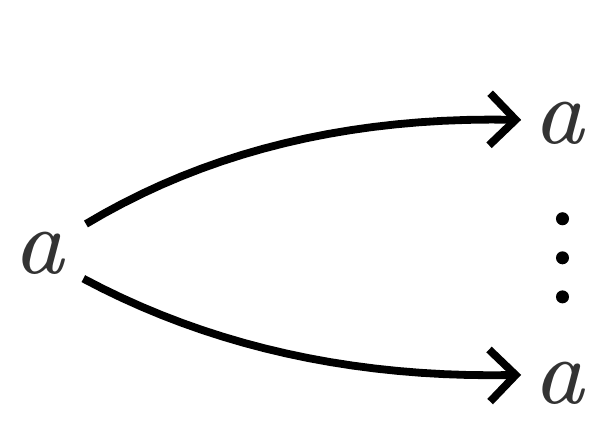}
  \end{array}
  $$
  \caption{"Detour" reduction rules}
  \label{fig:detours}
\end{figure}

\begin{figure}
  \captionsetup[subfigure]{justification=centering}
  \centering
  Let $\Gamma = A_1, \ldots, A_n$.
  \vspace{2em}
  
  \begin{subfigure}[b]{\textwidth}
    $$
    \begin{array}{r@{\qquad\mapsto\qquad}l}
      \vcenter{
        \prftree[r]{\rsf{var}}
          {\Gamma, x : A \vdash x : A}
      }
      &
      \includepdf{2.5cm}{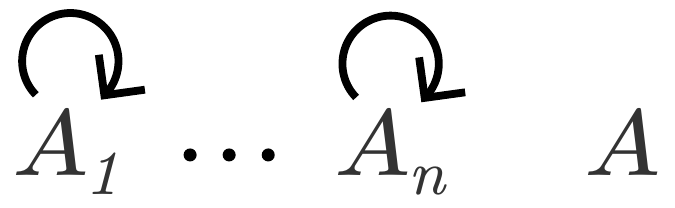}
      \\[2em]
      \vcenter{
        \prftree[r]{\rsf{abs}}
          {\Gamma, x : A \vdash t : B}
          {\Gamma \vdash \lambda x. t : A \to B}
      }
      &
      \includepdf{4.75cm}{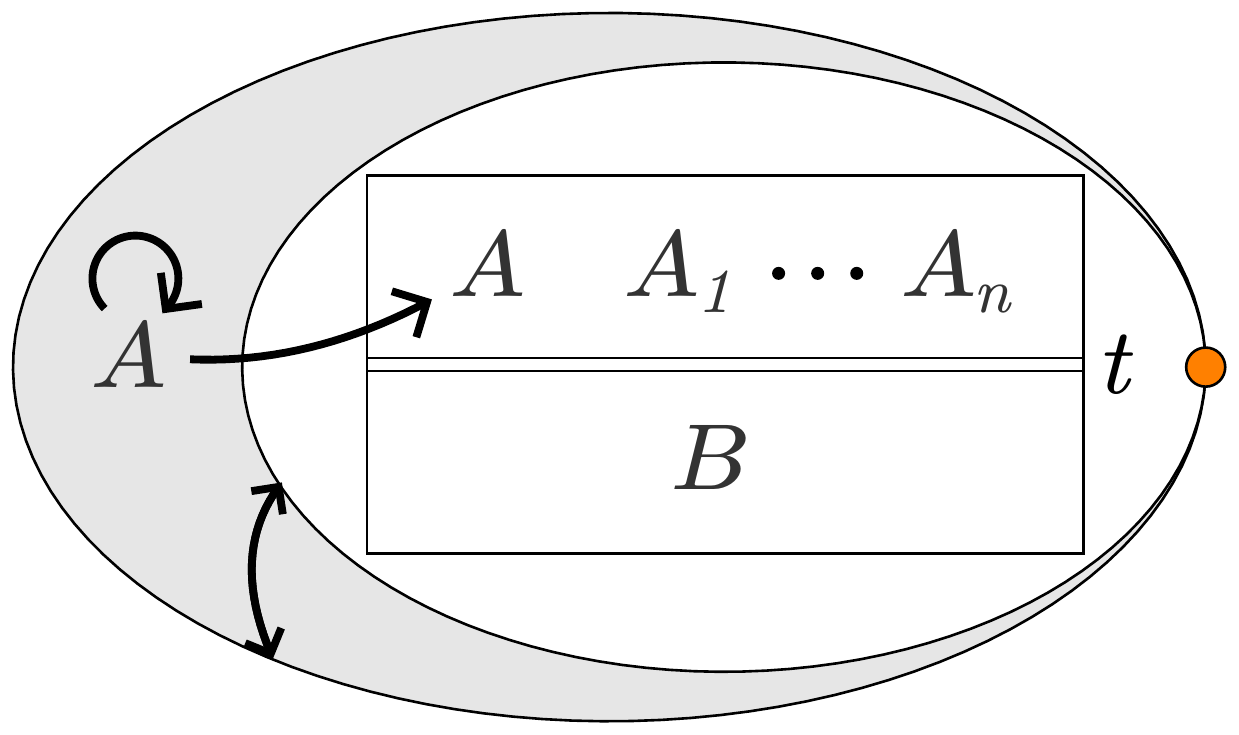}
      \\[2cm]
      \vcenter{
        \prftree[r]{\rsf{app}}
          {\Gamma \vdash t : A \to B}
          {\Gamma \vdash u : A}
          {\Gamma \vdash (t)u : B}
      }
      &
      \includepdf{4.75cm}{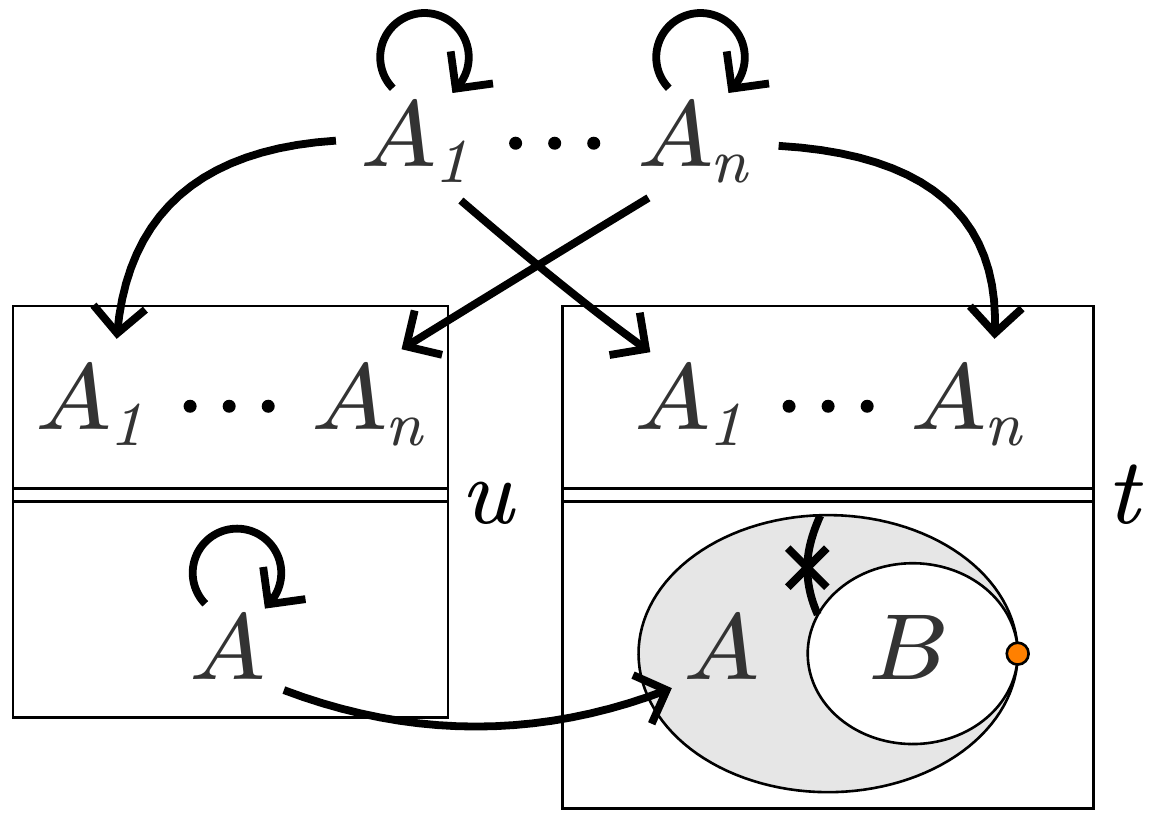}
    \end{array}
    $$
    \caption{Typing rules}
    \label{fig:simul-typing-rules}
  \end{subfigure}

  \vspace{2em}

  \begin{subfigure}[b]{\textwidth}
    \scalebox{0.9}{
    $
    \begin{array}{rclrcl}
      &&
      \vcenter{
        \prftree[r]{\rsf{app}}
          {\prftree[r]{\rsf{abs}}
            {\Gamma, x : A \vdash t : B}
            {\Gamma \vdash \lambda x. t : A \to B}}
          {\Gamma \vdash u : A~~~~\,}
          {\Gamma \vdash (\lambda x. t) u : B}
      }

      &&& \includepdf{6.4cm}{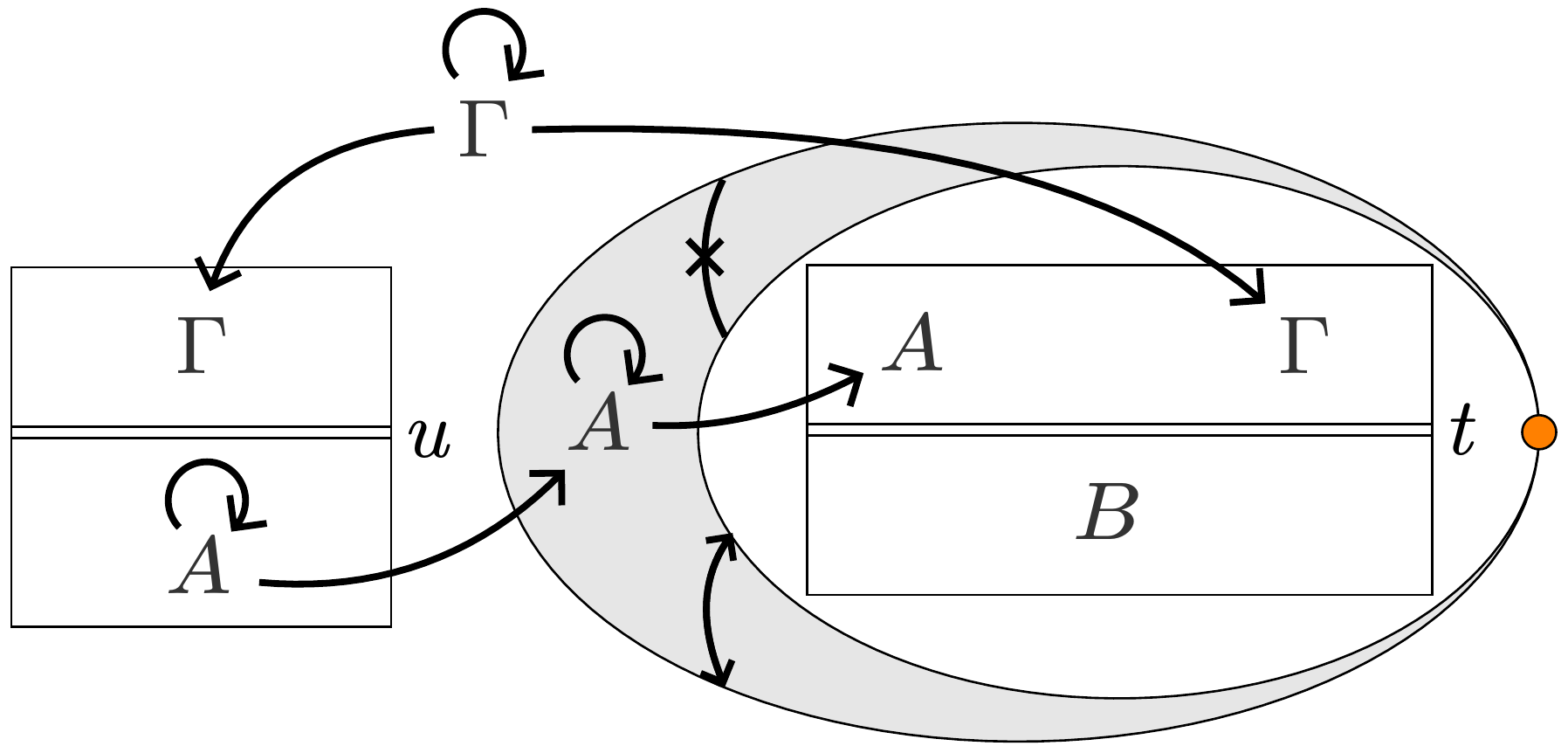} \\

      &&&&\red{aa}& \includepdf{6.4cm}{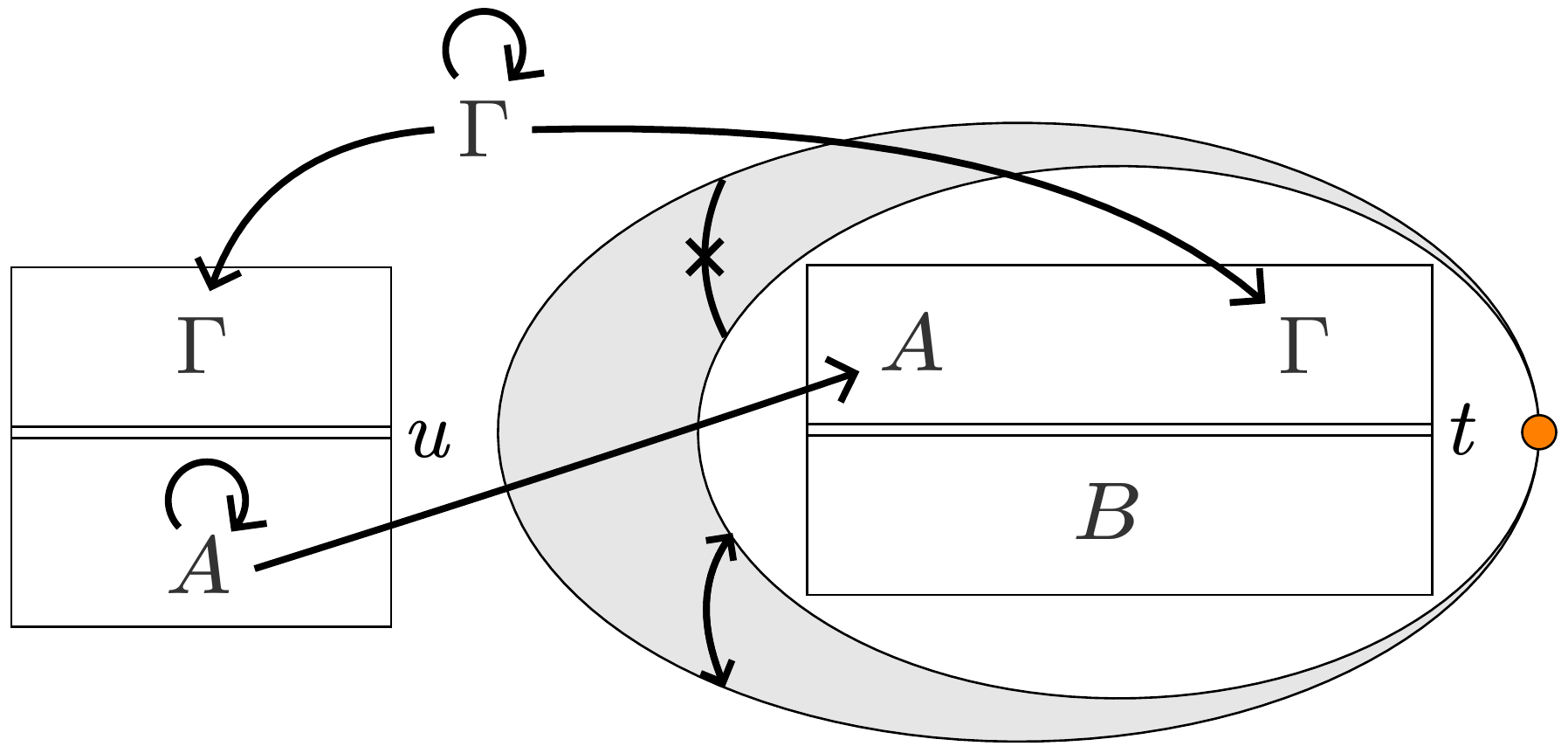} \\

      &\red{\beta}&
      \Gamma \vdash t\{u/x\} : B
      
      &&\red{ii}& \includepdf{6.4cm}{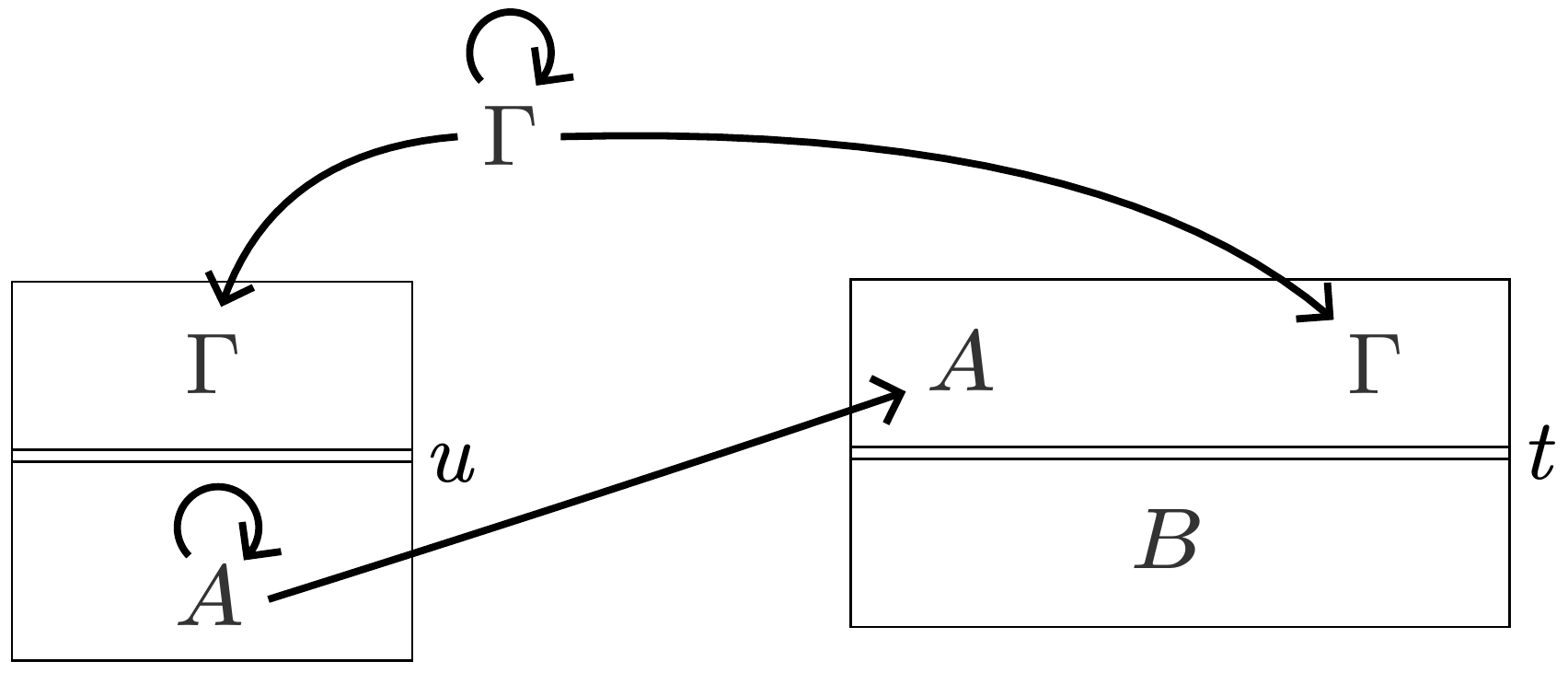}
    \end{array}
    $
    }
    \caption{$\beta$-reduction}
    \label{fig:simul-beta-reduction}
  \end{subfigure}
  \caption{Simulation of simply typed $\lambda$-calculus}
  \label{fig:simul}
\end{figure}

\end{document}